\documentclass[a4paper,usenatbib]{mnras}
\usepackage[T1]{fontenc}
\usepackage{ae,aecompl}
\usepackage{graphicx}
\usepackage{hyperref}
\usepackage{amsmath}
\usepackage{amssymb}
\usepackage{mathtools}
\usepackage{bm}
\usepackage{cleveref}
\usepackage[dvipsnames]{xcolor}

\title[Modelling stresses in neutron stars] 
{Modelling strains and stresses in continuously stratified  rotating neutron stars} 

\author[Giliberti et al.]{
E. Giliberti$^{1,4}$\thanks{elia.giliberti@unimi.it},
G. Cambiotti$^{2}$, 
M. Antonelli$^{3}$ 
and P.M. Pizzochero$^{1,4}$ 
\\
$^{1}$Dipartimento di Fisica, Universit\`a degli Studi di Milano, Via Celoria 16, 20133,Milano, Italy\\
$^{2}$Dipartimento di Scienze della Terra, Universit\`a degli Studi di Milano, Via Cicognara 7, Milano, 20129, Italy\\
$^{3}$Nicolaus Copernicus Astronomical Center, ul. Bartycka 18, 00-716 Warsaw, Poland\\
$^{4}$Istituto Nazionale di Fisica Nucleare, sezione di Milano, Via Celoria 16, 20133 Milano, Italy
}

\begin{document}
\pagerange{\pageref{firstpage}--\pageref{lastpage}} \pubyear{2018}
\maketitle
\label{firstpage}

\begin{abstract}

We introduce a Newtonian model for the deformations of a compressible and continuously stratified neutron star dropping the widely used Cowling approximation. 
 {The model is quite general and can be applied to a number of astrophysical scenarios as it allows to account for a great variety of loading forces.}
In this first analysis, the model is used to study the impact of a frozen adiabatic index in the calculation of rotation-induced deformations: we assume a polytropic equation of state for the matter at  equilibrium but, since chemical reactions may be slow, the perturbations with respect to the unstressed configuration are modeled by using a different adiabatic index. 
We quantify the impact of a departure of the adiabatic index from its equilibrium value on the stressed stellar configuration and we find that a small perturbation can cause large variations both in displacements and strains.
As a first practical application, we estimate the strain developed between two large glitches in the Vela pulsar  {showing that, starting from an initial unstressed configuration, it is not possible to reach the breaking threshold of the crust, namely to trigger a starquake. In this sense, the hypothesis that starquakes could  trigger  the unpinning of superfluid vortices is challenged} and, for the quake to be a possible trigger, the solid crust must never fully relax after a glitch, making the sequence of starquakes in a neutron star an history-dependent process. 

\end{abstract}


\section{Introduction}

The solid crust of a neutron star (NS) may develop stresses under the action of external loads, like the centrifugal force due to rotation \citep{link1998,fattoyev2018}, pinning of superfluid vortices to the crustal lattice \citep{rudermanI1991}, the presence of mountains \citep{usho2000,haskell2006} or intense  magnetic fields \citep{haskell_sam_2008}. 

The sudden setting of the crust in a strong gravitational field has been invoked in the modeling of many astronomical phenomena related to neutron stars, such as glitches \citep{ruderman1976} and flares \citep{blaes1989}, the possible neutron star precession \citep{pines1974nat,cutler2003} and the emission of gravitational waves \citep{laski2015PASA}, which may have direct consequences on the observed braking indexes of pulsars \citep{woan2018}. 

The stresses gradually build-up in the crust till a certain threshold, defined by the so-called breaking strain, is reached. At this point the elastic behaviour of the lattice abruptly ceases and a portion of the crust settles to a more relaxed configuration. 
This process is known as starquake (or crustquake) and its relation to glitches in pulsars \citep{baym1969,rudermanII1991} and to bursts in magnetars \citep{thompson1995,lander2015,keer2015MNRAS} is supported by different studies, underlining that the glitch sizes \citep{melatos2008,howitt2018ApJ}  and the burst-energy distribution seems to follow a power law \citep{cheng1996,gogus2000}, as do earthquakes on Earth. 
The quakes may also drive NSs precession, as explored by \citet{usho2000}, as well as the evolution of the magnetic field \citep{link1998, lander2019arXiv}. 
On the other hand, the rigid crust can sustain triaxial deformations (referred to as \emph{mountains} in the literature), which size may be enough to emit gravitational waves  detectable  in the near future \citep{ligo_first2017}. 

Despite such a large use of the crust failure hypothesis, there is still a lack of ``realistic'' and quantitative models for the study of crust deformations under different types of loading forces.
In fact, to date most of the studies rely on different approximations, either in the neutron star structure description, where the star is idealized as a uniform elastic sphere \citep[see e.g.][]{baym1969, franco2000, fattoyev2018}, a two-layers sphere \citep{giliberti2019PASA} or for using the Cowling approximation  
\citep[as in ][]{usho2000}, which is expected to  have a considerable impact on the estimates of the deformations and on the related quadrupole moment \citep{haskell2006}.

As a first step towards a better and more consistent description of elastic deformations in NSs, we adapt a model for deformations of the Earth \citep{sabadini_book} to describe neutron stars stresses due to a general loading force. Differently from other works, which are suitable for a specific kind of loading mechanism, the present approach allows to study the effect of a very general force (i.e. it can be easily adapted to model different astrophysical scenarios). 

To properly account for stratification, we allow for an arbitrarily large number of layers  with different values of the bulk and shear modulus, varying with continuity inside every layer \citep[see ][ for a comprehensive review of the properties of NSs crust]{chamel_livingreview}. 
At the boundaries between two layers the elastic properties of matter may be discontinuous. 

 {
The model presented is static, in the sense that we derive and solve the \emph{equilibrium} equations for the hydro-elastic configuration of a rotating NS. 
However, the stresses and the corresponding perturbations evolve on a certain typical timescale defined by the dynamics of the specific physical process under study. This is reflected in a non-adiabatic response of matter. 
For example, several studies on NS oscillations \citep{meltzer1966, cham1977, gourgo1995, HL2002} and thermal fluctuation in accreting neutron stars \citep{usho2000} pointed out that that the stellar crust matter can be not in full thermodynamical equilibrium during the build-up of internal stresses.
In particular, beta equilibrium is known to have an important role in the damping of NSs oscillations \citep{HL2002,yako2018,ander_ping2019}. 
In fact, chemical equilibrium is mediated by beta processes on highly neutron rich nuclei and such processes are expected to be extremely slow, so that the crustal layers can be out of beta equilibrium for a very long time, especially for cold (mature) neutron stars \citep{yako2001}.
Hence, the possibility of  a non-adiabatic response of matter may be relevant in the calculation of stresses in the crustal layers of a spinning-down (or spinning-up) neutron star. 
}

 {
To model the non-equilibrium response of matter the idea is the following. For a given equation of state (EoS), it is possible to define a related adiabatic index that determines the changes of pressure associated with variations of the local baryon density \citep{shapiro_book}. 
This fundamental quantity appears into the equations governing small-amplitude NSs pulsations \citep{campo1967}  and, in a relativistic context, provides a criteria of stability for cold stars \citep{meltzer1966, cham1977, gourgo1995}.
When the star is perturbed around the hydrostatic equilibrium configuration, the adiabatic index value must be calculated by taking into account for the possible slowness of the equilibration channels, typically mediated by the weak interaction \citep{haensel_book}. 
If the timescale of the perturbation is comparable or smaller than the ones of reactions, the adiabatic index is not simply the one defined by the equation of state at equilibrium as any change in density disturbs beta equilibrium.}
Therefore, we will consider a departure of the adiabatic index away from its equilibrium value and discuss how this affects the estimate of the stresses that can be built in a neutron star in which the  rotation period is slowly varying.  
 { As a first  astrophysical application we extend the analysis of stresses in spinning-down pulsars initiated in \cite{giliberti2019PASA} by means of incompressible models. However, following the recent analysis of \cite{fattoyev2018}, the present model will be used in a future work to discuss the breaking of the crust in spinning-up objects undergoing accretion and its implications on the maximum rotation rate of millisecond pulsars. }

The paper is organized as follows. 
In section 2 we introduce the main equations for compressible deformations of a neutron star. 
In section 3 we discuss how a departure of the adiabatic index from the value at chemical equilibrium, in the sense provided by \citet{HL2002}, can be expected during the gradual development of stresses.
In section 4 we conclude the description of the model by discussing the boundary conditions and writing the general elastic solution of the problem.
Section 5 is devoted to the numerical analysis of the displacements and strains induced by rotation  for a polytropic equation of state. 
Finally, in section 6 we calculate the strain developed by an isolated pulsar between two glitches 
 {and we compare the result with the one obtained with the simpler analytical model for incompressible pulsars of \cite{giliberti2019PASA}. }

\section{Main equations}

\begin{figure}
    \includegraphics[width=\columnwidth]{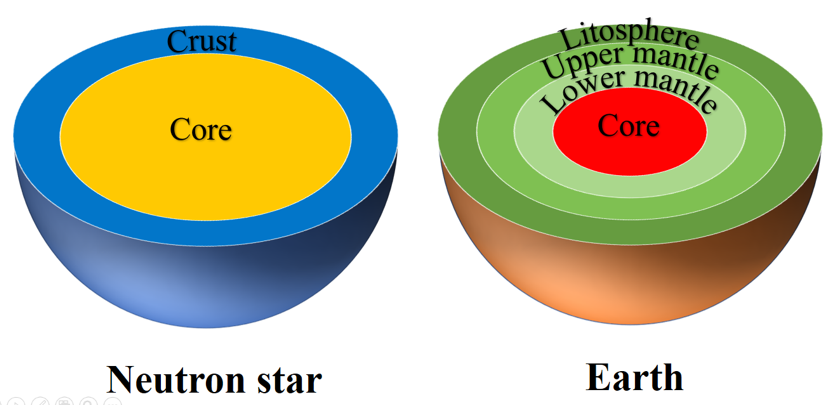}
    \caption{
    In the present study we adapt the $N$-layer model for the global elastic dynamics of the Earth described in \citet{sabadini_book} to describe a neutron star with a fluid core and an elastic crust.
    The density and the elastic coefficients vary continuously within each layer but may have discontinuities at the crust-core boundary.
    Since the model presented here is general, better comprehension of the elastic properties of the NS crust will allow to introduce additional layers, like a region containing the pasta phase at the bottom of the crust. }
    \label{fig:CONFRONTO NS TERRA}
\end{figure}

Our aim is to construct a framework to calculate the deformations of a compressible and self-gravitating neutron star  under the effect of specified forces in a Newtonian context. 
Building on the analogy with the Earth \citep{sabadini_book}, we describe the NS as an object with a fluid core topped by a number $N$ of stratified layers with different elastic properties, as sketched in Fig \ref{fig:CONFRONTO NS TERRA}. 
At the internal boundaries separating two layers of the NS, the material parameters may have step-like discontinuities due to the transition between specific phases of matter.

The approach developed here allows to consider stresses due to  {centrifugal forces, but also tidal and non-conservative forces, like vortex pinning to impurities in the crust  \citep{rudermanI1991}. In principle, it is also possible to include loads which account for inhomogeneities inside the star (the so-called bulk loads), as well as surface loads that could be used to model the effect of accretion of matter onto the crust.  However, it is known that surface loads need a specific technique to handle the boundary conditions \citep{sabadini_book}, so that we will not include them here to maintain the description of the model self-contained. }

We consider an unstressed and non rotating NS, characterized by the continuous density profile $\rho_0$, the pressure $P_0$ and the gravitational potential $\phi_0$. 
The momentum and Poisson equations are
\begin{align}
&
\boldsymbol{\nabla}P_0  + \rho_{0} \boldsymbol{\nabla}\phi_ 0  = 0
\label{EQ INIZIALE}
\\
&
\nabla^2 \phi_0 = 4\,\pi\,G\,\rho_0 \, ,
\label{POISSON INIZIALE}
\end{align}
which define the initial state of hydrostatic equilibrium. 
The displacement field $\boldsymbol{u}$ describes perturbations with respect to this spherically symmetric configuration and is defined as 
\begin{equation}
\boldsymbol{r} = \boldsymbol{x} + \boldsymbol{u}(\boldsymbol{x}) \, ,
\end{equation}
where $\boldsymbol{x}$  and $\boldsymbol{r}$ are the initial and the perturbed positions of the infinitesimal matter elements.
 {Hence, the total Cauchy stress tensor can be expressed} as\footnote{
In this work, as done previously in \cite{giliberti2019PASA}, we adopt the notation of \cite{sabadini_book}.  Given a generic quantity $f$, the \emph{local} increment $f^\Delta$ coincides with what is usually called Eulerian change \citep[see e.g. ][]{shapiro_book}.  The Lagrangian changes of $f$ are called \emph{material} increments and are $f^\delta$.
} 
\citep[cf. ][]{zdunik08}  
\begin{equation}
\boldsymbol{\tau}(\boldsymbol{r}) = -P_{0}(\boldsymbol{x})\,\boldsymbol{1} +  \boldsymbol{\tau}^\delta(\boldsymbol{x})  \,.
\end{equation}
 {In the above expression, $\boldsymbol{1}$ is the identity tensor and $\boldsymbol{\tau}^\delta$ is the material increment of the stress, which can be expressed by means of the usual Hooke's law \citep{love,landau_book}
\begin{equation}
\boldsymbol{\tau}^\delta 
= (\kappa-2/3\,\mu)\,(\boldsymbol{\nabla}\cdot\boldsymbol{u})\boldsymbol{1}
+
\mu\,\left(
\boldsymbol{\nabla}\boldsymbol{u}
+
(\boldsymbol{\nabla}\boldsymbol{u})^T
\right)
\label{DEF TAU} \, ,
\end{equation}
where $\kappa$ and $\mu$ are the bulk and shear moduli respectively. The transpose operation is indicated by the symbol $T$.} 

 {For a stressed configuration, the hydrostatic \eqref{EQ INIZIALE} and Poisson \eqref{POISSON INIZIALE} equations read}
\begin{align}
&
\boldsymbol{\nabla}\cdot\boldsymbol{\tau} - \rho\,\boldsymbol{\nabla}\Phi + \boldsymbol{h} = 0
\label{MOMENTUM}
\\
&
\nabla^2 \Phi = 4\,\pi\,G\,(\rho+\rho^t)-2\,\Omega^2
\label{POISSON EQUATION}
\end{align}
where $\boldsymbol{h}$ are the non-conservative body forces, $\rho$ is the mass density  of the deformed NS and $\rho^{t}$ the mass density of a possible companion. 
The potential $\Phi$ encodes all the conservative body forces and can be split as
\begin{equation}
\Phi = \phi+\phi^c+\phi^t \, ,
\label{definizione potenziale totale}
\end{equation}
where $\phi$ is the gravitational potential generated by the mass distribution of the NS, $\phi^t$ is the tidal potential due to the presence of the companion and $\phi^c$ is  the centrifugal potential. 
In particular, for a NS spinning on its axis with angular velocity $\Omega$  we have that 
$\nabla^2\phi^c = -2\,\Omega^2$.

Let us introduce the local incremental density  $\rho^\Delta$ and the local incremental potential $\Phi^\Delta$  as
\begin{align}
&
\rho(\boldsymbol{r}) = \rho_0(\boldsymbol{r}) +\rho^\Delta(\boldsymbol{r})
\label{ESPANSIONE RHO}
\\
&
\Phi(\boldsymbol{r}) = \Phi_0(\boldsymbol{r}) +\Phi^\Delta(\boldsymbol{r})
\label{ESPANSIONE PHI}
\end{align}
The first quantity is related to the displacement $\boldsymbol{u}$ via mass conservation
\begin{equation}
\rho^\Delta = - \boldsymbol{\nabla}\cdot(\rho_0\,\boldsymbol{u}), \,
\label{mass conservation}
\end{equation}
while the latter is given by
\begin{equation}
\Phi^{\Delta} 
= 
\phi^{\Delta}+\phi^{c}+\phi^{t} \, ,
\label{TERMINI POTENZIALE DELTA}
\end{equation}
where $\phi^{\Delta}$ is the incremental part of the gravitational potential generated only by the deformations of the NS.

 {We now use the definitions provided in Eqs \eqref{DEF TAU}, \eqref{ESPANSIONE RHO} and \eqref{ESPANSIONE PHI} and linearize  the momentum and Poisson equations. 
Using Eqs \eqref{EQ INIZIALE} and \eqref{POISSON INIZIALE} into the expansion of  Eqs \eqref{MOMENTUM} and \eqref{POISSON EQUATION}, immediately gives that the linearized  momentum and Poisson equations read \citep{sabadini_book} }
\begin{align}
&
\boldsymbol{\nabla}\cdot\boldsymbol{\tau}^\delta 
- 
\boldsymbol{\nabla}\cdot
\left(
\rho_0\,\boldsymbol{u}\cdot\boldsymbol{\nabla}\phi_0
\right)  
+
\boldsymbol{\nabla}\cdot(\rho_0\,\boldsymbol{u}) \, \boldsymbol{\nabla}\phi_0 
-  
\rho_0\,\boldsymbol{\nabla}\Phi^{\Delta}+\boldsymbol{h} 
= 0
\label{EQUAZIONE EQUILIBRIO GENERALE}
\\
&
\nabla^2 \Phi^{\Delta} = 4\,\pi\,G\,(-\boldsymbol{\nabla}\cdot(\rho_0\,\boldsymbol{u})+\rho^{t})
-2\,\Omega^2 \, .
\label{POISSON ESPANSA}
\end{align}
At this point it is convenient to work with the  usual spherical coordinate system $\left\{ r,\theta,\varphi\right\}$, that are the radial distance from the center of the NS, the colatitude and the longitude of a point respectively. 
Clearly, in the initial hydrostatic equilibrium configuration the elastic parameters $\kappa$ and $\mu$ are functions of the coordinate $r$ only. Furthermore, the gradient $\boldsymbol{\nabla}\phi_0 $ in the above equations can be expressed in terms of the  gravity acceleration for a spherically symmetric body,
\begin{equation}
g(r) = \partial_{r}\phi_{0}(r) = \frac{4 \pi G }{r^2} \int_0^r \rho_0( r') r'^2 dr' \, .
\label{finalmente}
\end{equation}
To find the stressed configuration we expand the potential $\Phi^{\Delta}$ in terms of spherical harmonics as in Eq \eqref{eq:philm}, see Appendix A for the conventions used. 
The same is done for the spheroidal and toroidal components of the displacement $\boldsymbol{u}$, defined in Eq \eqref{eq:uTS}. 
With some amount of algebra, Eq \eqref{EQUAZIONE EQUILIBRIO GENERALE} can be rearranged as

\begin{multline}
-\rho_{0}\partial_{r}\Phi_{\ell m}-\rho_{0}\partial_{r}\left(gU_{\ell m}\right)+\rho_{0}g\chi_{\ell m}+
\\
\partial_{r}\left[\left(\kappa-\frac{2}{3}\mu\right)\chi_{\ell m}+2\mu\partial_{r}U_{\ell m}\right]+
\\
\frac{1}{r^{2}}\mu\left[4r\partial_{r}U_{\ell m}-4U_{\ell m}+\ell\left(\ell+1\right)\left(3V_{\ell m}-U_{\ell m}-r\partial_{r}V_{\ell m}\right)\right]+
\\
h_{\ell m}^{R}=0
\label{eq:Momento radiale}
\end{multline}

\begin{multline}
-\frac{\rho_{0}}{r}\Phi_{\ell m}-\frac{\rho_{0}}{r}gU_{\ell m}+\frac{\left(\kappa-\frac{2}{3}\mu\right)}{r}\chi_{\ell m}+
\\
\partial_{r}\left[\mu\left(\partial_{r}V_{\ell m}+\frac{1}{r}U_{\ell m}-\frac{1}{r}V_{\ell m}\right)\right]+
\\
\frac{1}{r^{2}}\mu\left[5U_{\ell m}+3r\partial_{r}V_{\ell m}-V_{\ell m}-2\ell\left(\ell+1\right)V_{\ell m}\right]+
\\h_{\ell m}^{S}=0,
\label{eq:Momento tangenziale}
\end{multline}

\begin{multline}
\partial_{r}\left[ \mu \, \partial_{r}W_{\ell m} - \frac{\mu \, W_{\ell m}}{r} \right]
+
 \frac{3 \mu}{r}\partial_{r}W_{\ell m}-
\\
\frac{1+\ell\left(\ell+1\right)}{r^{2}}\mu W_{\ell m}  +h_{\ell m}^{T}=0
\, .
\label{eq:Momento toroidale}
\end{multline} 
In the above equations $U_{\ell m}$, $V_{\ell m}$ and $W_{\ell m}$ are the coefficients of the expansion in spherical harmonics relative to the  radial, tangential and toroidal parts of the displacement field $\boldsymbol{u}$, defined in Eqs \eqref{eq:disp_harminics1} and \eqref{eq:disp_harminics2}.  
Similarly, $h_{\ell m}^R$, $h_{\ell m}^S$, $h_{\ell m}^T$ are the expansion coefficients of the non-conservative forces, as given in Eq \eqref{eq:A13}. 
Finally, the scalars $\chi_{\ell m}$ are linked to the volume change according to
\begin{equation}
\boldsymbol{\nabla}\cdot\boldsymbol{u}=\sum_{\ell=0}^{\infty}\sum_{m=-\ell}^{\ell}\chi_{\ell m}Y_{\ell m}
\, .
\label{EQUAZIONE DELTA}
\end{equation}
The radial \eqref{eq:Momento radiale} and tangential \eqref{eq:Momento tangenziale} components  of the equilibrium equations are called \emph{spheroidal equations} while \eqref{eq:Momento toroidale} is called \emph{toroidal equation}.  
Following a similar procedure, the Poisson equation \eqref{POISSON ESPANSA} becomes
\begin{equation}
\nabla_{r}^{2}\Phi_{\ell m}
=
-4\pi G\left(\rho_{0}\chi_{\ell m}+U_{\ell m}\partial_{r}\rho_{0}\right)+4\pi G\rho_{\ell m}^{t}
\, ,
\label{POISSON DEFINITIVA}
\end{equation}
where
\begin{equation}
\nabla_{r}^{2}
=
\partial_{r}+\frac{2}{r}\partial_{r}-\frac{\ell\left(\ell+1\right)}{r^{2}}
\end{equation}
and $\rho_{\ell m}^{t}$ are the spherical harmonics coefficients relative to the mass density of the companion.

Equations (\ref{eq:Momento radiale}, \ref{eq:Momento tangenziale}, \ref{eq:Momento toroidale}, \ref{POISSON DEFINITIVA}) hold only for $\ell>0$; the case $\ell=0$ needs a specific treatment, as shown in detail in Appendix C. 
Furthermore, it is useful to notice that the toroidal equation \eqref{eq:Momento toroidale} is decoupled from Eqs (\ref{eq:Momento radiale}, \ref{eq:Momento tangenziale}, \ref{POISSON DEFINITIVA}). 

 {We now stick to a more specific physical scenario in which the deformations of the NS are due only to the fact that the star is rotating. This allows us to neglect the toroidal equation, which considerably simplifies the problem \citep{sabadini_book}. }
Neglecting the toroidal equation, Eqs  (\ref{eq:Momento radiale}, \ref{eq:Momento tangenziale}, \ref{POISSON DEFINITIVA}), are  second order differential equations in $U_{\ell m}$, $V_{\ell m}$ and $\Phi_{\ell m}$. Hence, it is convenient to recast them  into six differential equations of the first order introducing a vector  $\boldsymbol{y}_{\ell m}$ with six components, namely
\begin{equation}
\boldsymbol{y}_{\ell m}
=
\left(U_{\ell m},V_{\ell m},R_{\ell m},S_{\ell m},\Phi_{\ell m},Q_{\ell m}\right)^{T}
\, .
\label{VETTORE SFEROIDALE}
\end{equation}
Following \cite{sabadini_book}, we will refer to the components of the vector $\boldsymbol{y}_{\ell m}$ as \emph{spheroidal vector solutions}.
As shown in Appendix A, the meaning of the components is as follows: $U_{\ell m}$  and $V_{\ell m}$ are the coefficients of the expansion related to the radial and tangential displacements, $R_{\ell m}$  and $S_{\ell m}$ are related to the radial and tangential stresses,   $\Phi_{\ell m}$ describe the perturbation of the potential. 
Finally, in \eqref{VETTORE SFEROIDALE} the \emph{potential stress}  coefficients $Q_{\ell m}$ are defined as
\begin{equation}
Q_{\ell m}
=
\partial_{r}\Phi_{\ell m}+\frac{\ell+1}{r}\Phi_{\ell m}+4\pi G\rho_{0}U_{\ell m}
\, .
\label{POTENTIAL STRESS}
\end{equation}
Thanks to \eqref{VETTORE SFEROIDALE}, the whole system of equations (\ref{eq:Momento radiale}, \ref{eq:Momento tangenziale}, \ref{POISSON DEFINITIVA}) can be written in the more compact form
\begin{equation}
\frac{d\boldsymbol{y}_{\ell m}}{dr}
=
\boldsymbol{A}_{\ell}\left(r\right)\boldsymbol{y}_{\ell m}\left(r\right)-\boldsymbol{h}_{\ell m}\left(r\right).
\label{EQUAZIONE FONDAMENTALE EQUILIBRIO}
\end{equation}
Here $\boldsymbol{A}_{\ell}$ is a $6\times6$ matrix that defines the elastic response of the NS in its unstressed configuration (its form is explicitly given in Appendix B). 
The non-homogeneous term $\boldsymbol{h}_{\ell m}$ in equation \eqref{EQUAZIONE FONDAMENTALE EQUILIBRIO} contains the contributions of the non-conservative forces and is given by
\begin{equation}
\boldsymbol{h}_{\ell m}=\left(0,0,h_{\ell m}^{R},h_{\ell m}^{S},0,0\right)^{T},
\end{equation}
where the meaning of the superscripts $S$ and $R$ should be clear by looking at the quantities introduced in the spheroidal vector \eqref{VETTORE SFEROIDALE}. 
In the following we will consider only the effect of rotation, so that the vector $\boldsymbol{h}$ can be set to zero.

\section{Modeling the response of matter}

Following the standard description for cold catalyzed matter in a NS interior, we consider a barotropic EoS of the kind $P(n_b)$, $\rho(n_b)$, where $n_b$ is the baryon density.
In particular, the pressure is
\begin{equation}
P\left(n_{b}\right)=n_{b}^{2}\frac{d}{dn_{b}} \frac{E\left(n_{b}\right)}{n_{b}}
\, ,
\end{equation}
where $E\left(n_{b}\right)$ the ground state energy density at chemical equilibrium composition \citep{haensel_book}. 
The equation of state for matter at chemical equilibrium is characterized by the adiabatic index $\gamma_{eq}$, defined as
\begin{equation}
\gamma_{eq}(n_{b})=\frac{n_{b}}{P}\frac{\partial P\left(n_{b}\right)}{\partial n_{b}}.
\label{GAMMA EQUILIBRIO}
\end{equation}
 {
The use of a barotropic EoS is suitable when pressure-density perturbations are very slow with respect to the typical timescales of the chemical  reactions that carry the system towards the full thermodynamic equilibrium. 
Therefore, $\gamma_{eq}$ regulates the pressure response of matter to perturbations when the typical timescale of the dynamical process considered is larger than that of all the relevant equilibration channels.  
The opposite limit, in which all the reactions are so slow that the chemical composition is ``frozen''  has been considered in the context of NS oscillations \citep{HL2002}. More generally, we expect that the elastic response of the star can be affected by the slowness of the chemical equilibration reaction when matter does not have enough time to reach the complete thermodynamic equilibrium in the meanwhile stresses build-up.
In particular, \citet{yako2001} estimate that the equilibration timescale involving modified URCA processes scales as $\sim (2 \, \text{months})/T_9^6$, where $T_9$ is the internal temperature in units of $10^9\,$K. 
Hence, due to the strong dependence on the temperature of the star, the rotation-induced stresses that can develop on the timescale of years in a mature spinning-down (or spinning-up) NS should be calculated by taking into account that the adiabatic index for perturbations of matter may differ from $\gamma_{eq}$. 
}
If this is the case, the effective adiabatic index $\gamma_{f}$ that regulates the pressure-density perturbations, where the subscript $f$ stands for \emph{frozen}, depends explicitly also on the chemical fractions $x_{i}$ for the i-species, namely 
\begin{equation}
\gamma_{f}(n_{b},x_{i})
=
\frac{n_{b}}{P}\frac{\partial P\left(n_{b},x_{i}\right)}{\partial n_{b}} \, ,
\label{GAMMA FROZEN}
\end{equation} 
where the derivative is performed at fixed $x_i$ values \citep{HL2002}. 

It is useful to introduce also a third notion of  adiabatic index, that is derived directly from the  radial profiles of the thermodynamic quantities:  for a given spherically symmetric NS configuration, we define  a related \emph{effective} adiabatic index as
\begin{equation}
\gamma
=
\frac{\rho_0}{P_0 } \frac{\partial_r  P_0 }{\partial_r  \rho_0 }
=
-\frac{\rho_0^{2}  \, g }{P_0  \, \partial_{r}\rho_0  }  \, ,
\label{gamma effettivo}
\end{equation}
where $g$ is the modulus of the gravity acceleration for the spherical NS configuration considered.  

For a given barotropic  EoS  the  equilibrium radial profiles $P_0(r)$ and $\rho_0(r)$ can be obtained via Eqs \eqref{EQ INIZIALE} and \eqref{POISSON INIZIALE}. In this configuration the pressure-density relation supporting the star is characterized by the equilibrium adiabatic-index and $\gamma=\gamma_{eq}$ locally at every radius $r$. 
This ceases to be true if the EoS employed depends explicitly also on the chemical composition, as in Eq \eqref{GAMMA FROZEN}: in this case the relation 
\begin{equation}
\partial_r \rho_0 = \dfrac{\partial  \,  \rho}{\partial P} \, \partial_r P_0
\end{equation}
is no more valid\footnote{
Instead, we should consider 
$ \dfrac{\partial  \rho_0}{\partial r}=  \dfrac{\partial  \rho}{\partial P}   \dfrac{\partial  P_0}{\partial r}  + \dfrac{\partial  \rho}{\partial x_i}  \dfrac{\partial  x_{i0}}{\partial r}$.
}, so that $\gamma$ differs from both $\gamma_f$ and $\gamma_{eq}$. Hence, the  the effective adiabatic index can be a useful tool to probe a departure from the barotropic configuration due to a non-equilibrium stratification of matter \citep{Cambiotti2010}. 

\subsection{Polytropic equation of state}

To maintain consistence with the previous analysis of \cite{giliberti2019PASA}, we use a polytropic EoS with polytropic index $n=1$, namely
\begin{equation}
P\left(\rho\right)=K\rho^{2}=Km_{n}^{2}n_{b}^{2}
\, ,
\label{RELAZIONE POLITROPICA}
\end{equation}
Clearly, in this case the adiabatic index is not a function of $n_b$, 
but takes the constant value $\gamma_{eq}=(n+1)/n=2$. 
On the other hand, we have little clues about the actual value of $\gamma_{f}$. It is expected that  $\gamma_{f}>\gamma_{eq}$, but the actual relation between them strongly depends on the microscopic model underlying the specific EoS \citep{meltzer1966, cham1977, HL2002,usho2000}.
However, from the practical point of view, the uncertain value of $\gamma_f$ is not a strong limitation in the present work: we will assume different values and study how the estimated stress and strain change, starting from values that differ by only some percent from $\gamma_{eq}$, up to the incompressible limit  $\gamma_{f} \gg \gamma_{eq}$. 
Since   the main difference between the equilibrium and the frozen adiabatic index is expected to be in the crust \citep{usho2000}, we will vary the value of $\gamma_f$ only there, keeping the constant value of two in the core. 
Finally, for any realistic EoS, both $\gamma_{eq}$ and $\gamma_{f}$ have a complex dependence on the local properties of matter \citep[see e.g.][]{douchin2001,HL2002}. However, since the equilibrium adiabatic index  is just a constant, we will assume a constant $\gamma_{f}$ in the elastic layer as well.

\subsection{Strain angle and Tresca criterion }

The strain tensor is obtained from the displacement field $\boldsymbol{u}$ via \citep{love,landau_book}
\begin{equation}
\sigma_{ij}
=
\frac{1}{2}\left(\frac{\partial u_{i}}{\partial x_{j}}+\frac{\partial u_{j}}{\partial x_{i}}\right)
\, .
\end{equation}
To study the breaking of the elastic crust a failure criterion is needed. We assume the widely used  Tresca failure criterion \citep{failure_book}.
This criterion is based on the \emph{strain angle}, a local quantity $\alpha(r,\theta)$ that is the difference between the maximum and minimum eigenvalues of the strain tensor at a specific point. The   criterion assumes that, locally, the elastic behaviour of a material ceases when the strain angle approaches a particular threshold value $\sigma^{Max}$ known as the \emph{breaking strain},  
\begin{equation}
\alpha\approx\frac{1}{2}\sigma^{Max}
\, .
\label{TRESCA CRITERION}
\end{equation}
To date only order-of-magnitude estimates exist for the breaking strain; hence, the particular failure criterion assumed is of secondary importance\footnote{
	An alternative could be the Von Mises criterion, adopted e.g. by \citet{lander2015} and \citet{usho2000}.
	}.
The molecular dynamics simulations performed by \citet{horo2009} suggest that $\sigma^{Max} \sim 10^{-1}$ for a drop of nuclear matter. 
Using a completely different approach, \citet{baiko18} recently found that the maximum strain for  a polycrystalline crust is $\sim0.04$.
On the other hand, \citet{rudermanII1991} reasoned that, if crust has already undergone many cracks events, a macroscopic estimate for $\sigma^{Max}$ should be in the range $10^{-5} \div 10^{-3}$.
Hence, to take into account for the large uncertainties on the breaking strain, we will consider constant values of $\sigma^{Max}$ in the whole range $10^{-5}\div10^{-1}$.

\section{Boundary conditions}

To solve Eq \eqref{EQUAZIONE FONDAMENTALE EQUILIBRIO} we have to impose some boundary conditions. 
 {In the unstressed configuration,  the material parameters, consisting of the initial mass density $\rho_0$, the bulk modulus $\kappa$ and the shear modulus $\mu$ are continuous functions of the radial distance from the NS center $r$ within each layer.
In this first analysis, we assume that matter does not cross the interface between two elastic layers, meaning that material elements do not undergo phase transitions.} 
With these premises, it is known that all the spheroidal vector solutions in \eqref{VETTORE SFEROIDALE} are continuous across the boundaries between different layers \citep{sabadini_book}, say at $r=r_j$ for $j=1,...,N$, so that 
\begin{equation}
\boldsymbol{y}_{\ell m}\left(r_{j}^{-}\right)=\boldsymbol{y}_{\ell m}\left(r_{j}^{+}\right)  
\, .
\label{CONTINUITA' SFEROIDALE}
\end{equation}
This continuity requirement gives us a straightforward way to impose the boundary conditions at the surface of the NS and at the core-crust boundary,  {as explained in the following subsections}.

\subsection{Surface-vacuum boundary}

We can impose three simple conditions that have to be fulfilled by the spheroidal solution $\boldsymbol{y}$ at $r=a$, where $a$ is the stellar radius. 
The first is that the potential stress \eqref{POTENTIAL STRESS} must be continuous across the interface at $r=a$, 
\begin{equation}
Q_{\ell m}\left(a^{-}\right)=Q_{\ell m}\left(a^{+}\right)
\, .
\label{CONDIZIONE SU Q}
\end{equation}
To implement the condition \eqref{CONDIZIONE SU Q} in our model, let us first expand the centrifugal potential as
\begin{equation}
\phi^{c}\left(r,\theta,\varphi\right)
=
\phi_{00}^{c}\left(r\right)Y_{00}\left(\theta,\varphi\right)+\sum_{m=-2}^{2}\phi_{2m}^{c}\left(r\right)Y_{2m}\left(\theta,\varphi\right)
\, ,
\label{ESPANSIONE POTENZIALE CENTRIFUGO}
\end{equation}
where
\begin{equation}
\phi_{00}^{c}\left(r\right)=-\frac{\Omega^{2}r^{2}}{3}
\, 
\end{equation}
and
\begin{equation}
\phi_{2m}^{c}\left(r\right)
=
\frac{\Omega^{2}r^{2}}{3}\frac{\left(2-m\right)!}{\left(2+m\right)!}
\, Y^*_{2m}\left(\theta,\varphi\right)
\, .
\end{equation}
Note that the other harmonic coefficients of the expansion are zero, i.e. $\phi_{\ell m}^{c}=0$ for $\ell \neq 0,2$. Neglecting for the moment the $\ell=0$ term, we can assume 
\begin{equation}
\phi_{\ell m}^{c}\left(r\right)
=
\phi_{\ell m}^{c}\left(a\right)\left(\frac{r}{a}\right)^{\ell}
\, .
\label{ESPANSIONE CENTRIFUGO l>0}
\end{equation} 
The expansion of the gravitational and tidal potential is easy, since the Poisson equation \eqref{POISSON ESPANSA} reduces to the Laplace equation in the region between the NS and the body exerting the tidal force, placed at radius $a^{t}$. 
By imposing the regularity conditions for $r\rightarrow\infty$ and $r\rightarrow0$, we obtain
\begin{align}
& \phi_{\ell m}^{\Delta}\left(r\right)=\phi_{\ell m}^{\Delta}\left(a\right)\left(\frac{r}{a}\right)^{-\ell-1}
&\quad &
r>a
\label{ESPANSIONE POT GRAV}
\\
& \phi_{\ell m}^{t}\left(r\right)=\phi_{\ell m}^{t}\left(a\right)\left(\frac{r}{a}\right)^{\ell}
& &
r<a^{t} \, ,
\label{ESPANSIONE POT TID}
\end{align}
that, together with Eq \eqref{ESPANSIONE CENTRIFUGO l>0}, allow us to rewrite Eq \eqref{CONDIZIONE SU Q} as
\begin{equation}
Q_{\ell m}\left(a^{-}\right)
=
\frac{2\ell+1}{a}\left[\phi_{\ell m}^{c}\left(a^{+}\right)+\phi_{\ell m}^{t}\left(a^{+}\right)\right]
\, .
\label{CONDIZIONE FINALE SU Q}
\end{equation}
Besides Eq \eqref{CONDIZIONE FINALE SU Q}, we also impose that the tangential stress $S_{\ell m}$ must be zero in vacuum,
\begin{equation}
S_{\ell m}\left(a^{+}\right)=0
\, . 
\label{CONDIZIONE S R=a}
\end{equation} 
The same is valid for the radial stress $R_{\ell m}$ since the pressure outside the star is zero,
\begin{equation}
R_{\ell m}\left(a^{-}\right)=0
\, .
\label{CONDIZIONE R R=a}
\end{equation}
Finally, to find the final elastic solution, that will be discussed in section \ref{sec:solution}, it is useful to rearrange the three conditions (\ref{CONDIZIONE FINALE SU Q}, \ref{CONDIZIONE S R=a}, \ref{CONDIZIONE R R=a})  in the compact form 
\begin{equation}
\boldsymbol{P}_{1}\boldsymbol{y}\left(a^{-}\right)=\boldsymbol{b}
\label{condizioni di bordo superficie}
\end{equation}
where $\boldsymbol{P}_{1}$ is  {the} projector that selects only the third, fourth and sixth components of the spheroidal vector $\boldsymbol{y}$ and the vector $\boldsymbol{b}$ is defined as
\begin{equation}
\boldsymbol{b}=
\left(\begin{array}{c}
0\\
0\\
-\frac{\left(2\ell+1\right)}{a} \left(\phi_{\ell m}^{c}+\phi_{\ell m}^{t}\right)
\end{array}\right)
\, .
\label{DEFINIZIONE B}
\end{equation}

\subsection{Core-Crust boundary}

In our model the core is fluid and inviscid, so that it cannot support deviatoric stresses. 
Again, across the core-crust boundary  we can use the continuity of the spheroidal vector $\boldsymbol{y}$ but, differently with respect to what has been done in the previous subsection, we have to allow for a free slip of the material at the interface (the fluid core can slip under the crust). This request implies that 
\begin{equation}
\boldsymbol{y}(r_c^{+})
=
\left(\begin{array}{c}
U_{\ell m}\left(r_c^{-}\right)\\
0\\
R_{\ell m}\left(r_c^{-}\right)\\
0\\
\Phi_{\ell m}\left(r_c^{-}\right)\\
Q_{\ell m}\left(r_c^{-}\right)
\end{array}\right) \, + \, \left(\begin{array}{c}
0\\
C_{2}\\
0\\
0\\
0\\
0
\end{array}\right)
\, ,
\label{CRUST-CORE BOUNDARY}
\end{equation}
 where $r_c$ is the core-crust radius and $C_{2}$ is a constant of integration describing the tangential displacement. 

Now, the spheroidal vector solution in the core can be  found by setting $\mu=0$ and omitting the terms related to the loadings.
The first step is to rearrange Eqs (\ref{eq:Momento radiale}) and (\ref{eq:Momento tangenziale}) as 
\begin{align}
&
\frac{\partial_{r}R_{\ell m}}{\rho_{0}}-\partial_{r}\left(gU_{\ell m}\right)+g\chi_{\ell m}-\partial_{r}\Phi_{\ell m}=0
\label{FLUIDO RADIALE}
\\
&
\frac{R_{\ell m}}{\rho_{0}}-gU_{\ell m}-\Phi_{\ell m}=0
\, .
\label{FLUIDO TANGENZIALE}
\end{align}
Inserting these two equations into the Poisson one (\ref{POISSON DEFINITIVA}), we obtain
\begin{equation}
\nabla_{r}^{2}\Phi_{\ell m}=4\pi G\partial_{r}\rho_{0}\frac{\Phi_{\ell m}}{g}
\, .
\end{equation}
Since $\partial_{r}\rho_{0}=0$ must hold at the center of the star, the regularity of the potential in $r=0$ implies
\begin{equation}
\lim_{r\rightarrow0}r^{-\ell}\psi_{\ell m}\left(r\right)=1 
\, ,
\end{equation}
with $\Phi_{\ell m}\left(r\right)=C_{1}\psi_{\ell m}\left(r\right)$.

To proceed further, we subtract the radial derivative of Eq \eqref{FLUIDO TANGENZIALE} from \eqref{FLUIDO RADIALE} and use the relation
\begin{equation}
R_{\ell m}=\kappa\chi_{\ell m},
\label{frizzo}
\end{equation}
which is valid in the fluid limit, to obtain the so-called Adams-Williamson relation, 
\begin{equation}
\frac{\kappa}{\rho_{0}^{2}}\left(\partial_{r}\rho_{0}+\frac{\rho_{0}^{2}g}{\gamma P}\right)\chi_{\ell m}=0
\, .
\label{ADAMS-WILLIAMSON}
\end{equation}
For our purposes, it is convenient to rearrange the above equation  as  \citep{Cambiotti2010, Cambiotti2013}
\begin{equation} 
\frac{\kappa}{\rho_{0}^{2}}\frac{\partial_{r}\rho_{0}}{\gamma}\left(\gamma-\gamma_{eq}\right)\chi_{\ell m}=0 \, . 
\label{wa generale}
\end{equation}
Clearly, when $\gamma=\gamma_{eq}$  (i.e. the stratification is compressional, see section 3),  the above equation is automatically satisfied. 
In particular, the Adams-Willliamson relation is satisfied in the fluid core, so that the two equations \eqref{FLUIDO RADIALE} and \eqref{FLUIDO TANGENZIALE} are not linearly independent; this provides a way to constrain the radial stress  at the core-crust interface at $r=r_c$:
\begin{equation}
R_{\ell m}
=
\rho_{0} \, g \left[\, U_{\ell m} + \frac{\Phi_{\ell m}}{g}\right] = \rho_{0} \, g \, C_{3} \, ,
\end{equation}
where the constant  $C_3$ is the difference between the radial displacement $U_{\ell m}$ and the so-called geoid displacement
\begin{equation}
U^{geoid}_{\ell m}(r) =-\frac{\Phi_{\ell m}\left(r\right)}{g\left(r\right)}.
\label{GEOID DISPLACEMENT}
\end{equation}
evaluated at $r=r_c$  \citep{sabadini_book}.

Note that in the case of compressional stratification the volume change within the core is undetermined: below the core-crust interface we cannot specify the displacement and radial stresses with the above assumptions. However, this does not constitute a problem because we are interested only in the deformation of the crust, which is uniquely determined by the boundary conditions. 

The constants $C_{1}$, $C_{2}$ and $C_{3}$ define the elastic  solution at the core-crust boundary, that can be written as
\begin{equation}
y_{\ell m} \left(r_c \right)=\left(\begin{array}{c}
-C_{1} \, \frac{\psi_{\ell m}}{g}+C_{3}\\
C_{2}\\
\rho_{0}gC_{3}\\
0\\
C_{1}\psi_{\ell m}\\
C_{1}q_{\ell m}+4\pi G\rho_{0}C_{3}
\end{array}\right),
\label{666theNumberOfTheBeast}
\end{equation}
where we have defined 
\begin{equation}
q_{\ell m}=\partial_{r}\psi_{\ell m}+\frac{\ell+1}{r}\psi_{\ell m}-\frac{4\pi G}{g}\psi_{\ell m} \, .
\end{equation}
In view of Eq \eqref{666theNumberOfTheBeast}, the core-crust boundary conditions \eqref{CRUST-CORE BOUNDARY} are conveniently  arranged in the compact form
\begin{equation}
\boldsymbol{y}_{\ell m}\left(r_c^{+}\right)=\boldsymbol{I}_{C} \, \boldsymbol{C}
\, ,
\label{satana}
\end{equation}
where $\boldsymbol{I}_{C}$ is the $6\times3$ matrix 
\begin{equation}
\boldsymbol{I}_{C}=\left(\begin{array}{ccc}
-\psi_{\ell}\left(r_c\right)/g\left(r_c\right) & 0 & 1\\
0 & 1 & 0\\
0 & 0 & g\left(r_c\right)\rho_{0}\left(r_c^{-}\right)\\
0 & 0 & 0\\
\psi_{\ell}\left(r_c\right) & 0 & 0\\
q_{\ell}\left(r_c\right) & 0 & 4\pi G\rho_{0}\left(r_c^{-}\right)
\end{array}\right)
\end{equation}
and $\boldsymbol{C}$ is the 3-vector 
\begin{equation}
 \boldsymbol{C}=\left(C_{1},C_{2},C_{3}\right).
\end{equation}

\subsection{Elastic solution}
\label{sec:solution}

The general solution of the differential system \eqref{EQUAZIONE FONDAMENTALE EQUILIBRIO} reads
\begin{equation}
\boldsymbol{y}_{\ell m}\left(r\right)=\boldsymbol{\Pi}_{\ell}\left(r,r_{0}\right)\boldsymbol{y}_{\ell m}(r_{0})-\int_{r_{0}}^{r}\boldsymbol{\Pi}_{\ell}\left(r,r'\right)\boldsymbol{h}_{\ell m}\left(r'\right)dr'.
\label{GENERAL SOLUTION}
\end{equation}
The first terms on the right side of Eq \eqref{GENERAL SOLUTION} is the homogeneous solution, while the second term is a particular solution which accounts for the external non-conservative forces. 
Furthermore, the  propagator matrix $\boldsymbol{\Pi}_{\ell}$ solves the  homogeneous equation
\begin{equation}
\frac{d\boldsymbol{\Pi}_{\ell}\left(r,r'\right)}{dr}
=
\boldsymbol{A}_{\ell}\left(r\right)\boldsymbol{\Pi}_{\ell}\left(r,r'\right)
\, ,
\end{equation}
with the condition
\begin{equation}
\boldsymbol{\Pi}_{\ell}\left(r',r'\right) = \boldsymbol{1}
\, .
\end{equation}
At each boundary between the layers we have to impose the continuity of the propagator, namely
\begin{equation}
\boldsymbol{\Pi}_{\ell}\left(r_{j}^{+},r'\right)
=
\boldsymbol{\Pi}_{\ell}\left(r_{j}^{-},r'\right)
\, .
\end{equation}
If we choose the core-crust radius as the starting point of the integration in Eq \eqref{GENERAL SOLUTION}, namely  $r_{0}=r_c^{+}$, it is easy to use Eq \eqref{satana} to show that
\begin{equation}
\boldsymbol{y}_{\ell m}\left(r\right)
=
\boldsymbol{\Pi}_{\ell}\left(r,r_c^{+}\right)\boldsymbol{I}_{C}\boldsymbol{C}-\boldsymbol{w}\left(r\right)
\, ,
\label{equazione elasticità compatta}
\end{equation}
where  
\begin{equation}
\boldsymbol{w}\left(r\right)
=
\int_{r_c^{+}}^{r}\boldsymbol{\Pi}_{\ell}\left(r,r'\right)\boldsymbol{h}_{\ell m}\left(r'\right)dr'
\, .
\end{equation}
The three constants of integration in the vector $\boldsymbol{C}$ can be estimated by imposing the conditions at the star surface via Eq \eqref{condizioni di bordo superficie}, so that Eq \eqref{equazione elasticità compatta} now reads
\begin{multline}
\boldsymbol{y}_{\ell m}\left(r\right)
=
\boldsymbol{\Pi}_{\ell}\left(r,r_c\right)\boldsymbol{I}_{C}
\left[
\boldsymbol{P}_{1}\boldsymbol{\Pi}_{\ell}\left(a^{-},r_c^{+}\right)\boldsymbol{I}_{C}
\right]^{-1}
\\
\left(
\boldsymbol{P}_{1}\boldsymbol{w}\left(a^{-}\right)+\boldsymbol{b}
\right)
-
\boldsymbol{w}\left(r\right)
\, .
\label{ronaldo}
\end{multline}
This equation represents the  {most general} response of the star to the  {internal (such as vortex pinning), centrifugal and tidal} loads and it uniquely determines the spheroidal deformations and the potential within the  crust, as well as the radial and tangential spheroidal stresses and the potential stress. 

In our explicit case of study, the only external force is conservative (i.e. the centrifugal force) and the solution in Eq \eqref{ronaldo} assumes the simpler form
\begin{equation}
\boldsymbol{y}_{\ell m}\left(r\right)
=
\boldsymbol{\Pi}_{\ell}\left(r,r_c\right)\boldsymbol{I}_{C} 
\left[
\boldsymbol{P}_{1}\boldsymbol{\Pi}_{\ell}\left(a^{-},r_c^{+}\right)\boldsymbol{I}_{C}
\right]^{-1} \boldsymbol{b}
\, .
\end{equation}
Moreover, when the deformations with respect to the spherical reference configuration are induced only by rotation, the displacement field $\boldsymbol{u}$ is the sum of only two contributions arising from the zero and second harmonics\footnote{The fact that $\boldsymbol{u}$ is a sum of only the $\ell=0,2$ and $m=0$ harmonic contributions depends on the fact that we have assumed a constant rotation axis. Taking into account also for a possible nutation would require the contributions of other $m\neq0$ harmonics.}, namely $\boldsymbol{u}=\boldsymbol{u}_{00}+\boldsymbol{u}_{20}$. 

 {This section concludes the general description of the model. In the following we numerically study in details the deformations of rotating neutron stars.}

\section{Numerical solution for the polytrope {\lowercase{$n=1$}} }

As anticipated in section 3, we study the behaviour of a neutron star described by a polytrope with $n=1$. 
In particular, we are interested in comparing the displacements and strains of neutron stars with different masses and for different values of the frozen adiabatic index.

 {
The peculiar polytropic index $n=1$ has been chosen because it gives a degenerate mass-radius relation (the radius $a$ and the mass $M$ are independent of each other). 
Therefore, following the same approach outlined in \cite{giliberti2019PASA}, we can fix the pair $(a,M)$ with the mass and radius values obtained by using a realistic equation of state and solving the Tolman-Oppenheimer-Volkoff equations (TOV). 
The same can be done for the radius of the core-crust transition.
}

\begin{figure}
		\includegraphics[width=\columnwidth]{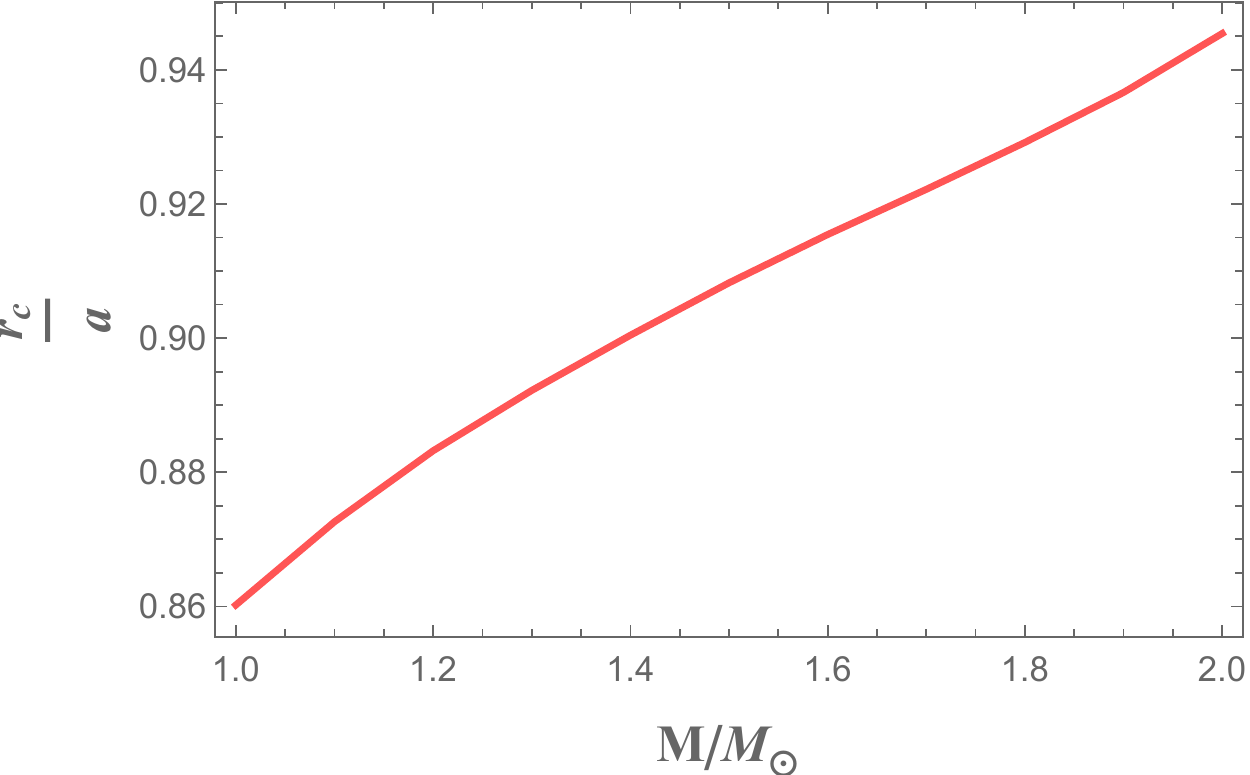}
            \caption{ Normalized core-crust radius $r_c/a$ as a function of the stellar mass $M$, obtained for the SLy EoS.}
    \label{fig:rc vs massa}
\end{figure}

 {
More explicitly,  we consider a realistic equation of state, say the SLy4 \citep{douchin2001}, solve the TOV equations, and use the obtained values of $a$ and $r_c$ to fix the mass and the radius of the polytropic configuration.
In this way, despite the fact that the radial density profile is the one obtained in a Newtonian framework by using the polytrope $n=1$ (that contains no information about the core-crust transition), in the following numerical analysis we still have reasonable values for the stellar radius $a$ and for the width $a-r_c$ of the elastic layer, see Fig \ref{fig:rc vs massa}. 
}

For a given radius $a$ and mass $M$ in the range $1 M_\odot \div 2 M_\odot $ we can calculate the corresponding value of $K$ in Eq \eqref{RELAZIONE POLITROPICA} and of the central density $\rho_{ce}=\rho(r=0)$ as 
\begin{equation}
 K=\frac{2}{\pi}a^{2}G
\qquad 
\rho_{ce}=\frac{M}{4\pi^{2}}\left(\frac{K}{2\pi G}\right)^{-3/2}.
\end{equation}
The crust-core transition is set at the fiducial density $1.5\times10^{14}\text{g/cm}^3 $, giving a core-crust transition at $r_c \approx 0.90 \, a$ for a standard neutron star with $M=1.4M_{\odot}$, as can be seen in  Fig \ref{fig:rc vs massa}. 

 {
	Now that the equilibrium structure of the NS has been fixed, we have to find the corresponding spheroidal solution $\boldsymbol{y}$ for a given angular velocity $\Omega$. 	In the following, if not otherwise stated, we show the results for a $1.4M_{\odot}$ NS, which characteristics are reported in Tab \ref{tab:table 1}.
}	

\begin{table}
\centering
	\caption{
	Parameters for the reference configuration of a neutron star used in the numerical analysis. 
	Since the rotation induced deformations and stresses are proportional to $\Omega$, the angular velocity is set to the value  $1$ rad/s so that the numerical results can be easily rescaled for different angular velocities.
	}
	\label{tab:table 1}
	\begin{tabular}{c|c|c|c|c}
		\hline
		$a$&  $\rho_{ce}$ & $\mu_{c}$ & $\Omega$ & $d(\Omega)$
		\\
		cm &  g/cm$^3$ &   dyn/cm$^2$ &  rad/s  &    
		\\
        \hline
        $1.17\times10^{6}$ & $1.38\times~10^{15}$ & $10^{30}$ & $1$ & $6.3\times 10^{-4}$
        \\
        \hline
\end{tabular}
\end{table}

 {
In the numerical calculations the physical quantities are normalized by using the stellar radius $a$, the central density $\rho_{ce}$, the shear modulus $\mu_{c}$ at the core-crust interface and the angular velocity $\Omega$  of the particular NS under consideration. The first two parameters vary with $M$, while the other two are fixed. In particular, the vector solution $\boldsymbol{y}$ is rescaled as
\begin{equation}
\boldsymbol{y}=d\left(\Omega\right)\times\left(\begin{array}{c}
a\\
a\\
\mu_{c}\\
\mu_{c}\\
v^{2}\\
v^{2}/a
\end{array}\right)\tilde{\boldsymbol{y}} \, ,
\end{equation} 
where the tilde superscript indicates the dimensionless quantities. 
An overall scaling is given by the dimensionless factor
\begin{equation}
d\left(\Omega\right)=\frac{1}{3}\frac{\Omega^{2}a^{2}}{v^{2}}
\, ,
\label{d}
\end{equation}
where the velocity $v$ is defined as $v=\sqrt{\mu_{c}/\rho_{ce}}$. 
The above ratio sets the relative importance of the centrifugal force with respect to the elastic reaction of matter and, for practical purposes, can be expressed as
%
%
\begin{multline}
d(\Omega)
=
6.3 \times 10^{-4} \left( \frac{ \Omega }{ 1 \text{rad/s} } \right)^2 
\left( \frac{ \rho_{ce} }{ 1.38 \times 10^{15} \text{g/cm}^3 } \right)
\times
\\
\times
\left( \frac{ \mu_c }{ 10^{30} \text{dyn/cm}^2 } \right)^{-1} \left( \frac{ a }{ 1.17 \times 10^6 \text{cm} } \right)^{2} ,
\end{multline}
calculated by using the parameters in Table 1 as a reference.}

 {
The elastic response of the star is fixed by the EoS and by the poorly known parameters $\kappa$ and $\mu$ in the crust.
For the shear modulus we follow the same prescription guessed by \citet{cutler2003}, namely
\begin{equation}
\mu(r) = 10^{-2} \times P(r)
\, .
\label{ANDAMENTO MU}
\end{equation}
On the other hand,  the elastic modulus $\kappa$ is linked to the adiabatic index by the relation
\begin{equation}
\kappa=\gamma_{eq, f} P \, ,
\label{LINK KAPPA GAMMA} 
\end{equation}
where the choice of $\gamma_{eq}$ or $\gamma_f$ depends on the sloweness of the loading mechanism.
}

To summarize, the stellar mass, together with the EoS, the adiabatic index governing perturbations, the shear modulus and the core-crust transition density completely fix the elastic behaviour of the star.
 {We can now study the effects of the centrifugal force, starting from a non-rotating and unstressed reference configuration.
}

\subsection{Slow dynamics}
\label{SEZIONE GAMMA EQ}

We set at first the equilibrium bulk modulus in \eqref{LINK KAPPA GAMMA}, namely
\begin{equation}
\kappa\left(r\right) = \gamma_{eq} P(r)  =  2 P(r)
\, .
\end{equation}
With this choice we are implicitly assuming that the rotational dynamics of the NS has a much longer timescale compared to the one of the chemical reactions near equilibrium.

Using the parameters given in Table \ref{tab:table 1}, the displacement with respect to the non-rotating configuration turns out to be $u_{r}(r=a, \theta=\pi/2)\simeq 4.2\times10^{-3}\,$cm at the equator. 

The centrifugal potential is particular as its expansion consists of only two  harmonic parts having $m=0$ and $l=0,2$. The one with degree $\ell=0$ gives a smaller contribution to the total displacement with respect to the $\ell=2$ contribution as we find that $2.5 \leq |U_{20}(r)/U_{00}(r)| \leq2.9$. 
Furthermore, the  $\ell=0$ contribution corresponds to a global increase of volume of the star since it is  positive everywhere.

\begin{figure}
		\includegraphics[width=\columnwidth]{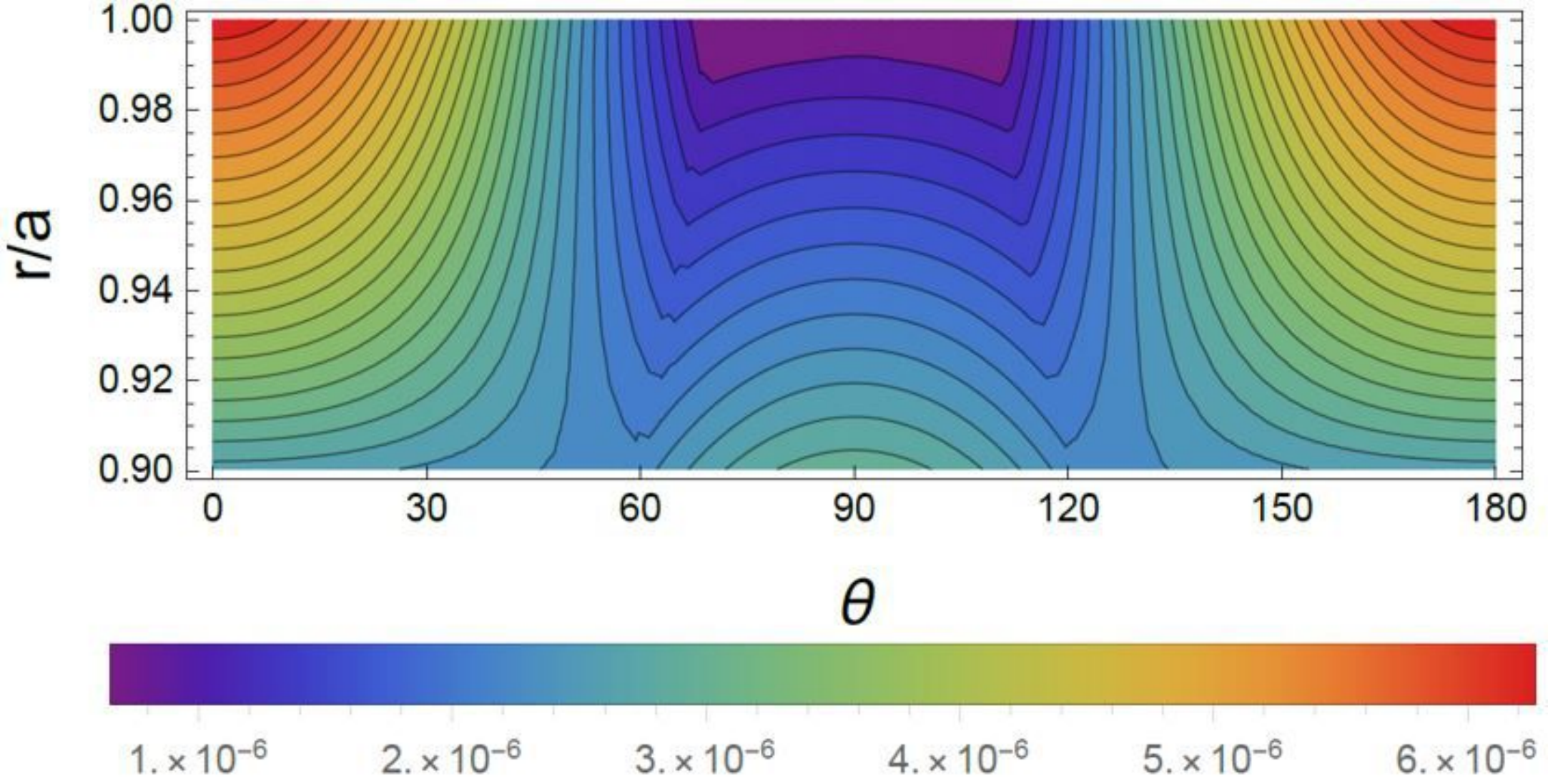}
    \caption{Color map of the normalized strain angle $\tilde{\alpha}$ as a function of
the colatitude  and of the normalized radius $r/a$. The region
shown here refers to the crustal layer, from $r = r_c$ to $r = a$ for a $M=1.4M_{\odot}$ NS. 
 {The  largest values are reached at the poles, near the stellar surface.}}
    \label{fig:Contour strain 1.4 gamma 2}
\end{figure}

We now explore the possibility of crust failure, by calculating the strain angle $\alpha$ and using the Tresca criterion in Eq \eqref{TRESCA CRITERION}. 
 {
In Fig \ref{fig:Contour strain 1.4 gamma 2}, the normalized strain angle $\tilde{\alpha}=\alpha/d(\Omega)$ is shown  as a function of the colatitude $\theta$ and of the normalized radius $r/a$.} 
Our analysis shows that:

\begin{enumerate}

\item Contrary to the uniform and incompressible model studied by \citet{franco2000} and \citet{fattoyev2018}, the strain angle is an increasing function of the radius. 
This happens because in the present model the shear modulus is not a constant but a decreasing function of the stellar radius, implying that the strain is expected to be larger near the surface, as was also noticed by  \citet{cutler2003}.

\item Differently from the incompressible and uniform model, where  $\alpha(r,\theta)$ is always peaked at the equator (as shown by \cite{giliberti2019PASA}), we find that the maximum value of the strain angle $\alpha^{Max}$ occurs at the poles. 

\item Concerning the dependence on the stellar mass, $\alpha^{Max}$ is a decreasing function of $M$, as can be seen in Fig \ref{fig:Strain vari masse gamma equilibrio}.  In this case it is not convenient to plot the normalized strain angle $\tilde{\alpha}$, since also the normalizing factor $d$ depends on the stellar mass through $a$ and $v$.
 {
This behaviour of $\alpha^{Max}(M)$  can be  understood by considering a very simplified description of a NS with only one uniform elastic layer extending from the center to the crust. In this case the displacement $\boldsymbol{u}$ turns out to be proportional to the dimensionless factor \citep{baym1971,link1998} }
\begin{equation}
W=\frac{\Omega^{2}a^{2}}{v_{K}^{2}} \, , 
\end{equation}
which is the ratio between the squared equatorial velocity $\Omega^{2}a^{2}$ and the Keplerian one $v_{K}=\sqrt{GM/a}$.  
The dependence of  $\alpha^{Max}$  on $M$ is almost entirely due to $W$, which is proportional to  
\begin{equation}
W \propto  \frac{a^{3}}{GM} \propto \frac{1}{\rho} \, ,
\end{equation}
where $\rho$ is the density of the uniform star \citep{giliberti2019PASA}.
Therefore, for more massive (i.e. denser) stars, smaller displacements are expected. 
This leads us to conclude that this reasonable behaviour, which could appear as a by-product of the simplified uniform model  of \citet{baym1971} is also a feature  of more refined models. 
\begin{figure}
		\includegraphics[width=\columnwidth]{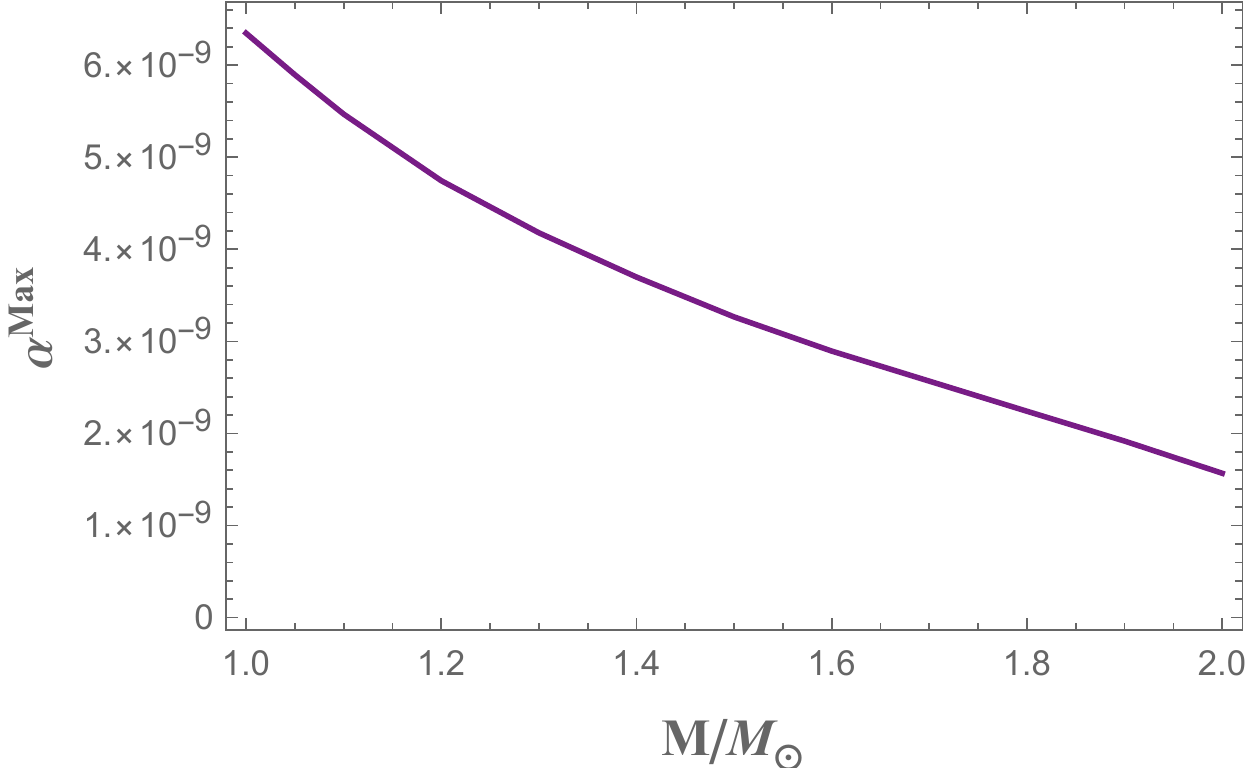}
    \caption{The maximum strain angle as a function of the mass, assuming $\gamma=\gamma_{eq}$ and $\Omega=1$ rad/s. }
    \label{fig:Strain vari masse gamma equilibrio}
\end{figure}

 
\end{enumerate}

\subsection{Fast dynamics}

We now investigate what happens when the dynamical timescale of the perturbations is fast with respect to the one of chemical reactions, so that
\begin{equation}
\kappa\left(r\right) = \gamma_{f} \, P(r)
\, .
\end{equation}
The polytropic EoS that we employ is purely phenomenological and does not carry any information about the value for the non-equilibrium adiabatic index. Therefore, we study the displacements, stresses and strains in three different cases, characterized by the adiabatic indexes $\gamma_f= 2.1$, $200$, $\infty$. For comparison purposes, we consider also the limiting case discussed in the previous section in which $\gamma_{eq}$ is used.
The value $\gamma_f=2.1$ is $5\%$ larger than the equilibrium adiabatic index, while the choice $\gamma_f=200$ is   just a numerical counterpart of the analytical incompressible limit given by $\gamma \rightarrow \infty$.

As discussed in section 3,  we vary only the crust adiabatic index, while the core maintains its equilibrium compressibility also in the limiting case in which the crust is treated as incompressible. 
Hence, also in the limiting case $\gamma_{f}=\infty$ the radial displacement is different from zero because the core modifies its shape during the spin-down, loading the crust. 

\begin{figure}
		\includegraphics[width=\columnwidth]{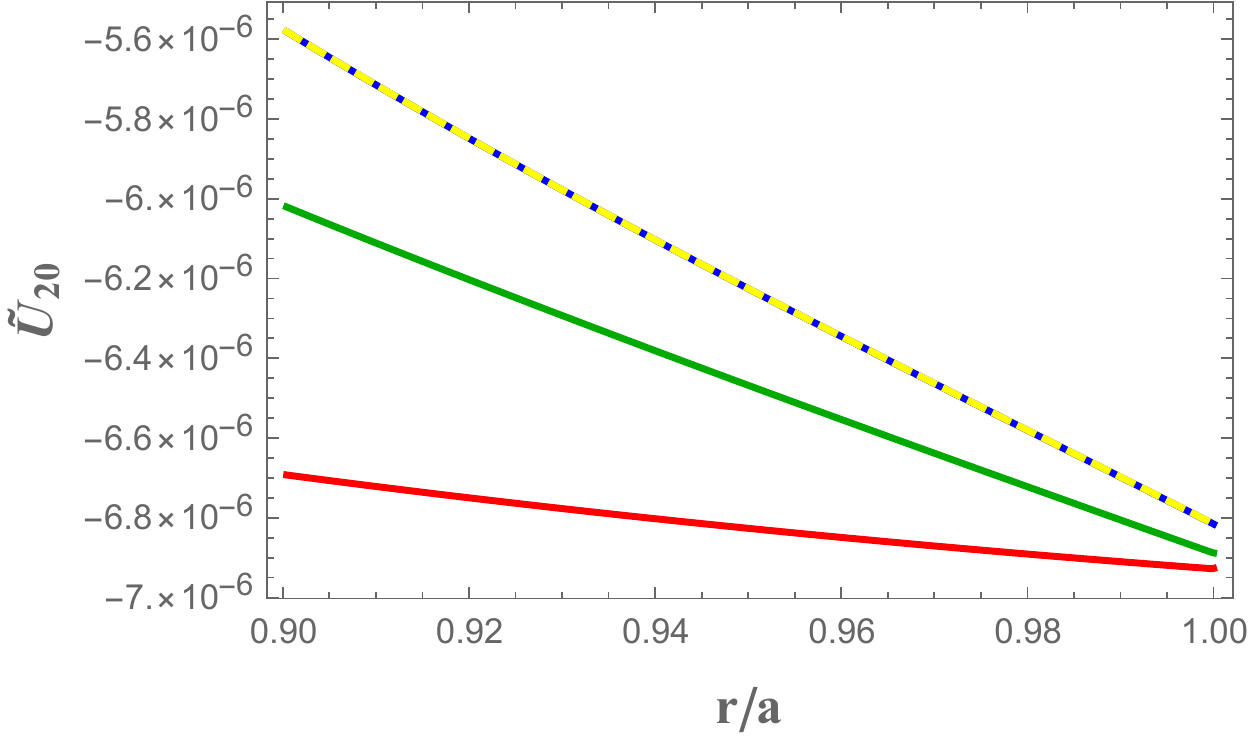}
		\includegraphics[width=\columnwidth]{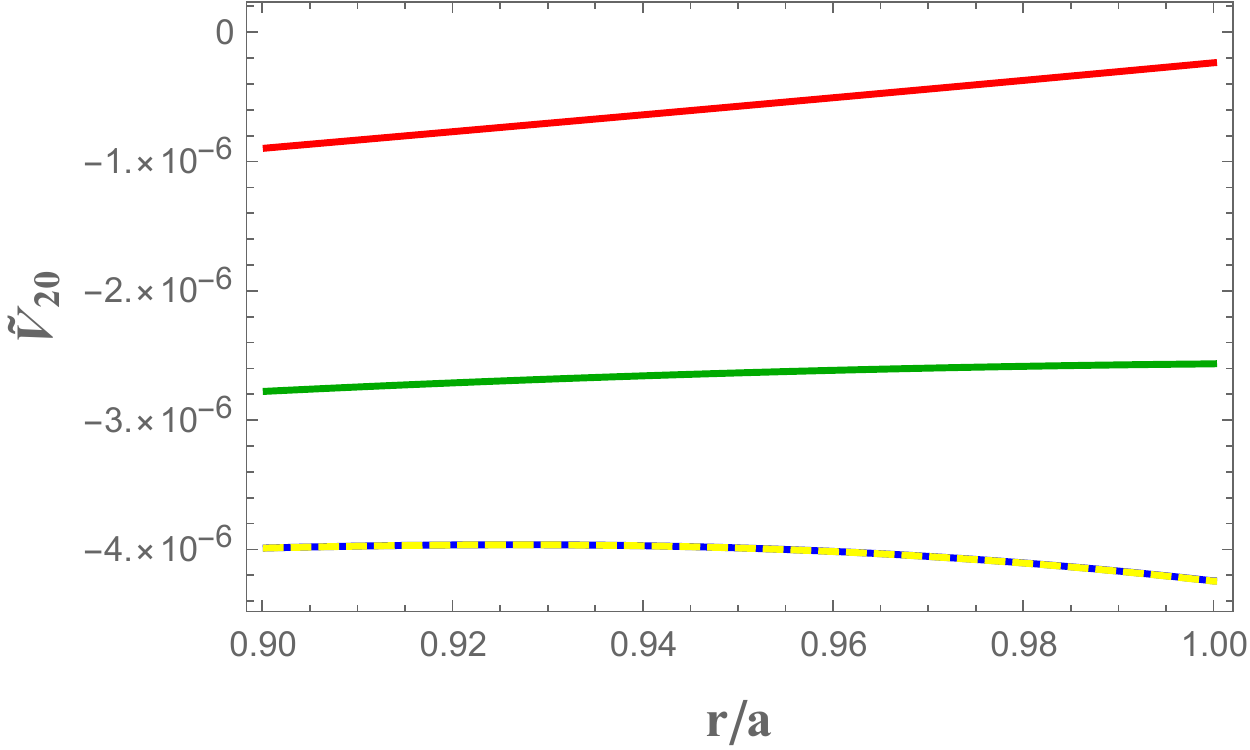}
\caption{
	Normalized $\tilde{U}_{20} $ (top) and $\tilde{V}_{20} $ (bottom) displacements as function of the normalized radius. The plot refers to the crustal region, extending from  $r=r_c$ to $r=a$. 
	Each curve corresponds to a different value of the assumed adiabatic index: $\gamma_f=\gamma_{eq}$ (red), $\gamma_f=2.1$ (green), $\gamma_f=200$ (blue) and $\gamma_f=\infty$ (yellow dashed).}
    \label{fig:U e V vari gamma}
\end{figure}

\begin{figure}
		\includegraphics[width=\columnwidth]{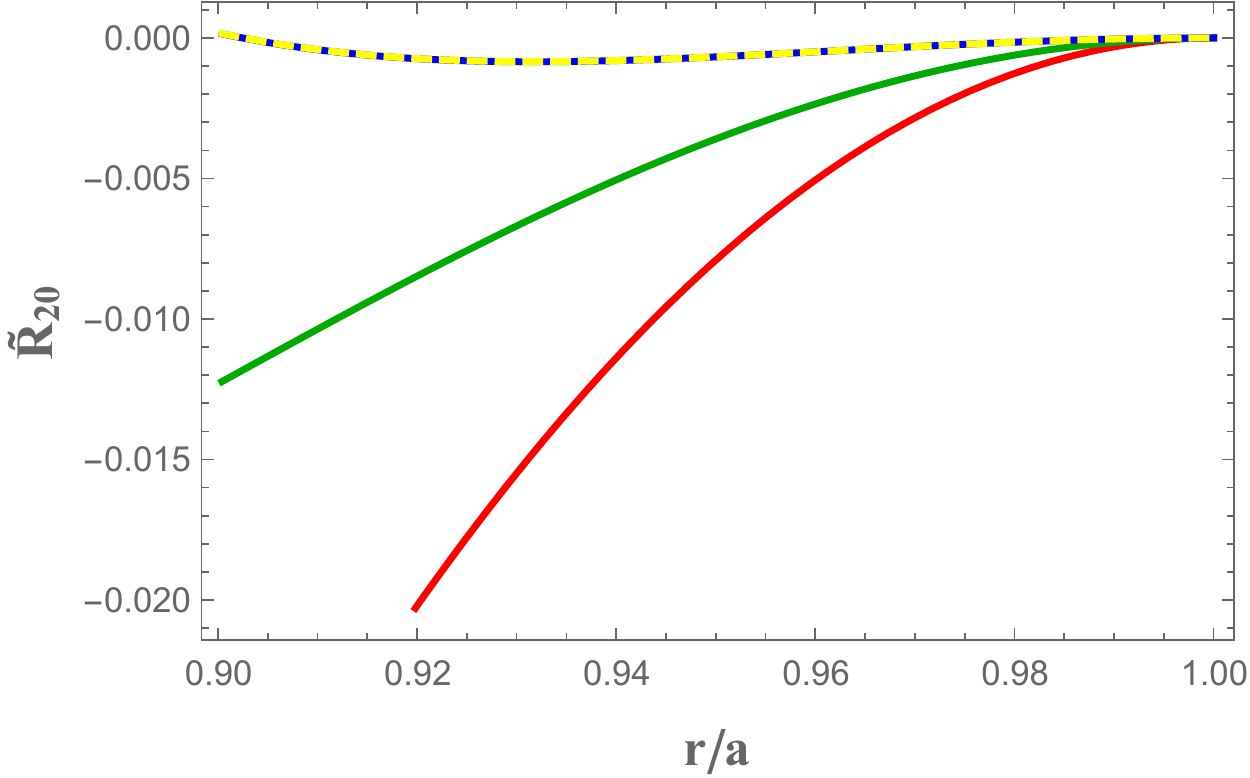}
		\includegraphics[width=\columnwidth]{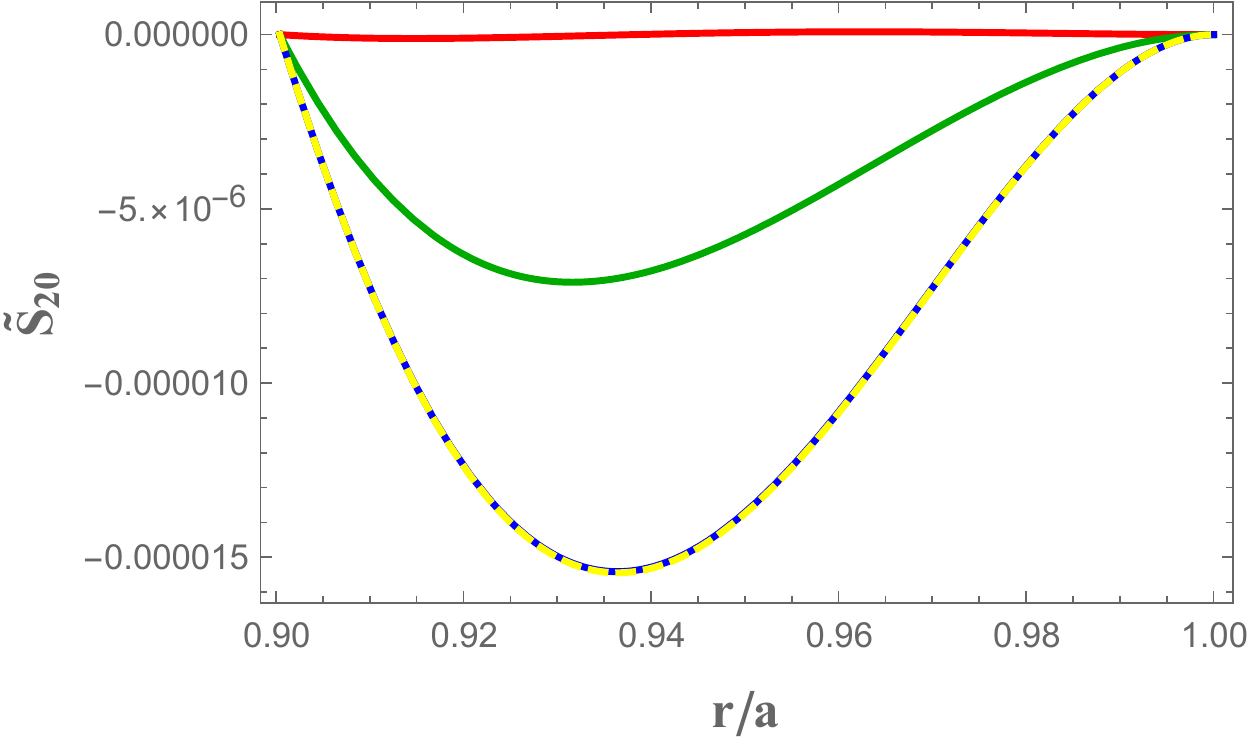}
\caption{
	Normalized $\tilde{R}_{20} $ (top) and $\tilde{S}_{20} $ (bottom) stresses as function of the normalized radius. The plot refers to the crustal region, extending from  $r=r_c$ to $r=a$. Each curve corresponds to a different value of the adiabatic index: $\gamma_f=\gamma_{eq}$ (red), $\gamma_f=2.1$ (green), $\gamma_f=200$ (blue) and $\gamma_f=\infty$ (yellow dashed).}
    \label{fig:R e S politropo vari gamma}
\end{figure}

 {The numerical solution of the model shows that even a small departure from the equilibrium value of the adiabatic index carries the system to a configuration similar to the incompressible one, as can be seen in Figs \ref{fig:U e V vari gamma} and \ref{fig:R e S politropo vari gamma}.  
In particular, Fig  \ref{fig:U e V vari gamma} shows the normalized values of $\tilde{U}_{20}$ and $\tilde{V}_{20}$ for a $M=1.4M_{\odot}$ rotating neutron star, according to the values listed in Tab \ref{tab:table 1}.
We notice that  the response of the star to the same change of the centrifugal force is quite different in the equilibrium scenario with respect to the frozen ones. The same consideration is valid for the radial stress $\tilde{R}_{20}$ and the tangential stress $\tilde{S}_{20}$, as shown in Fig \ref{fig:R e S politropo vari gamma}.
}

\begin{figure}
		\includegraphics[width=\columnwidth]{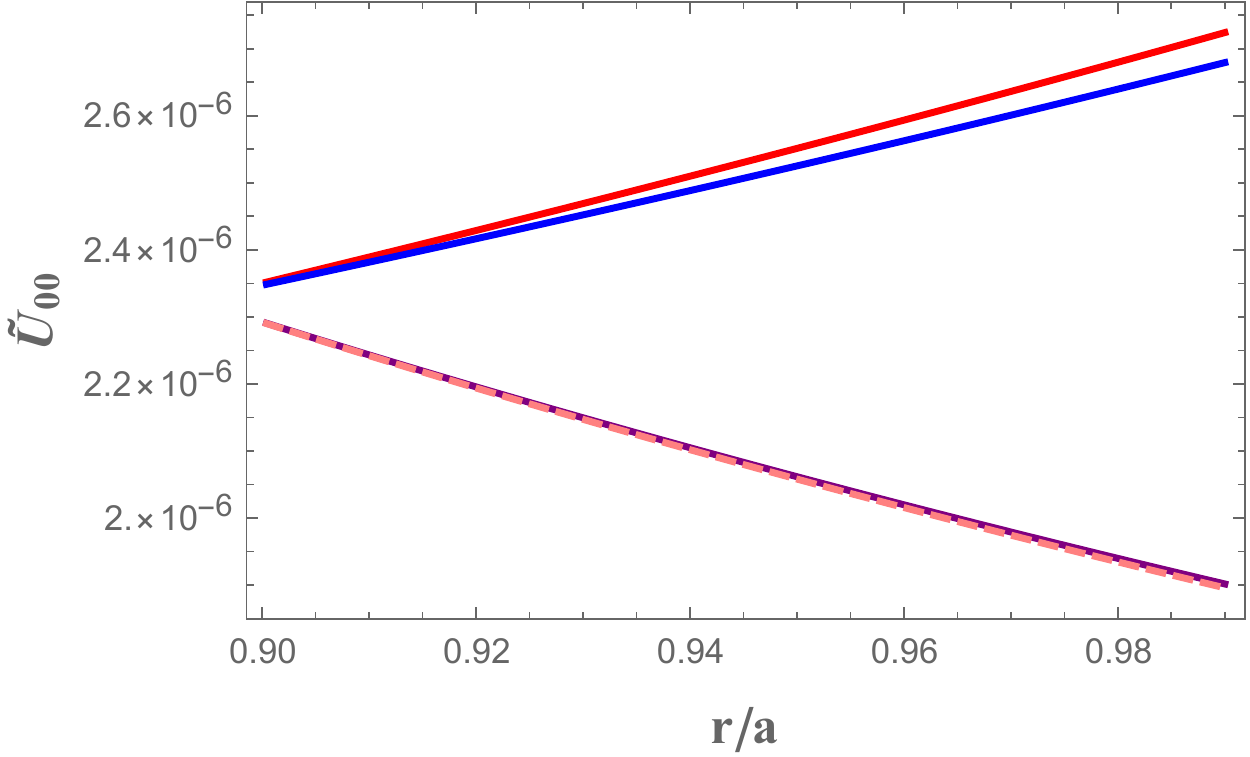}
\caption{
	Normalized $\tilde{U}_{00}$ displacement as function of the normalized radius, in the region between $r=r_c$ and $r=a$. As in the other figures, the curves refer to the results calculated by considering $\gamma_f=\gamma_{eq}$ (red), $\gamma_f=2.1$ (blue), $\gamma_f=200$ (purple) and $\gamma_f=\infty$ (pink, dashed). The curves for $\gamma_f=200,\infty$ are superimposed.}
    \label{fig:U0 vari gamma}
\end{figure}

 {For the same stellar configuration, Fig  \ref{fig:U0 vari gamma} represents the radial displacement $\tilde{U}_{00}$. In all cases we find that $\tilde{U}_{00}>0$, so that  the star  undergoes a global increase of volume due to the effect of rotation, as expected.
However, there is a marked difference between the equilibrium scenario and the incompressible limit, with the change of the slope of the plotted curve. 
}

\begin{figure}
		\includegraphics[width=\columnwidth]{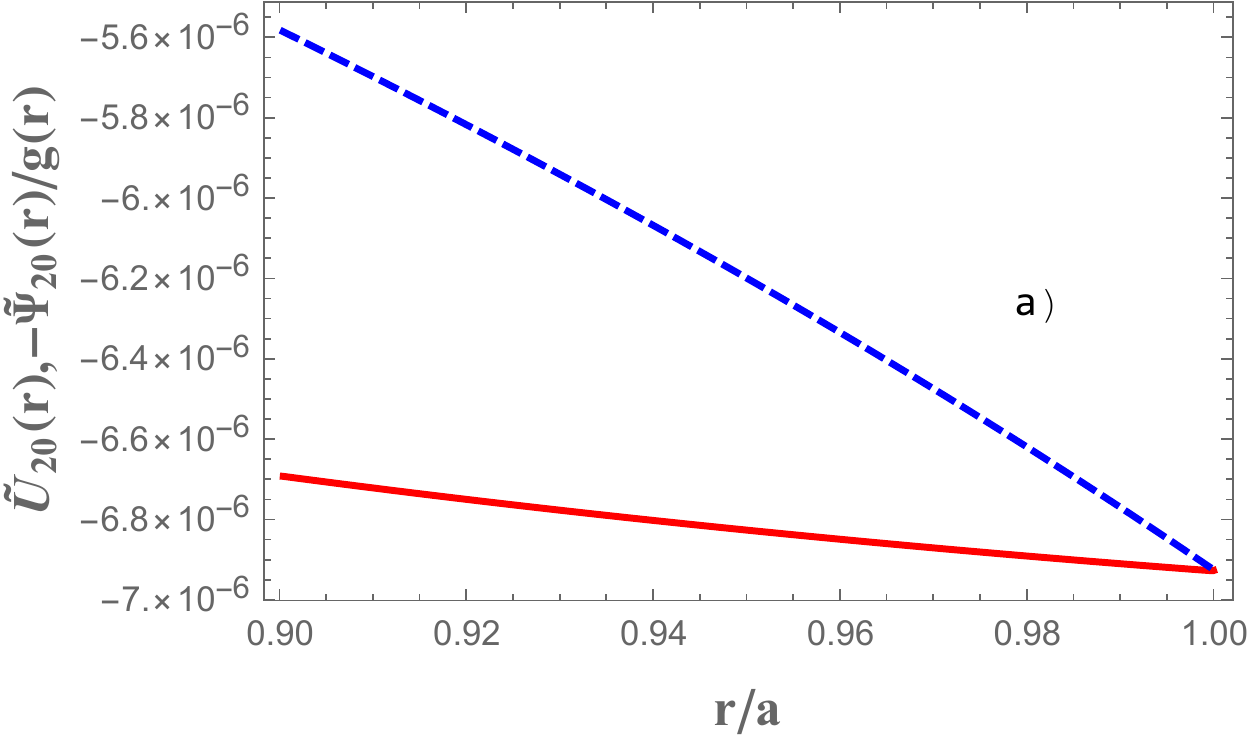}
        \includegraphics[width=\columnwidth]{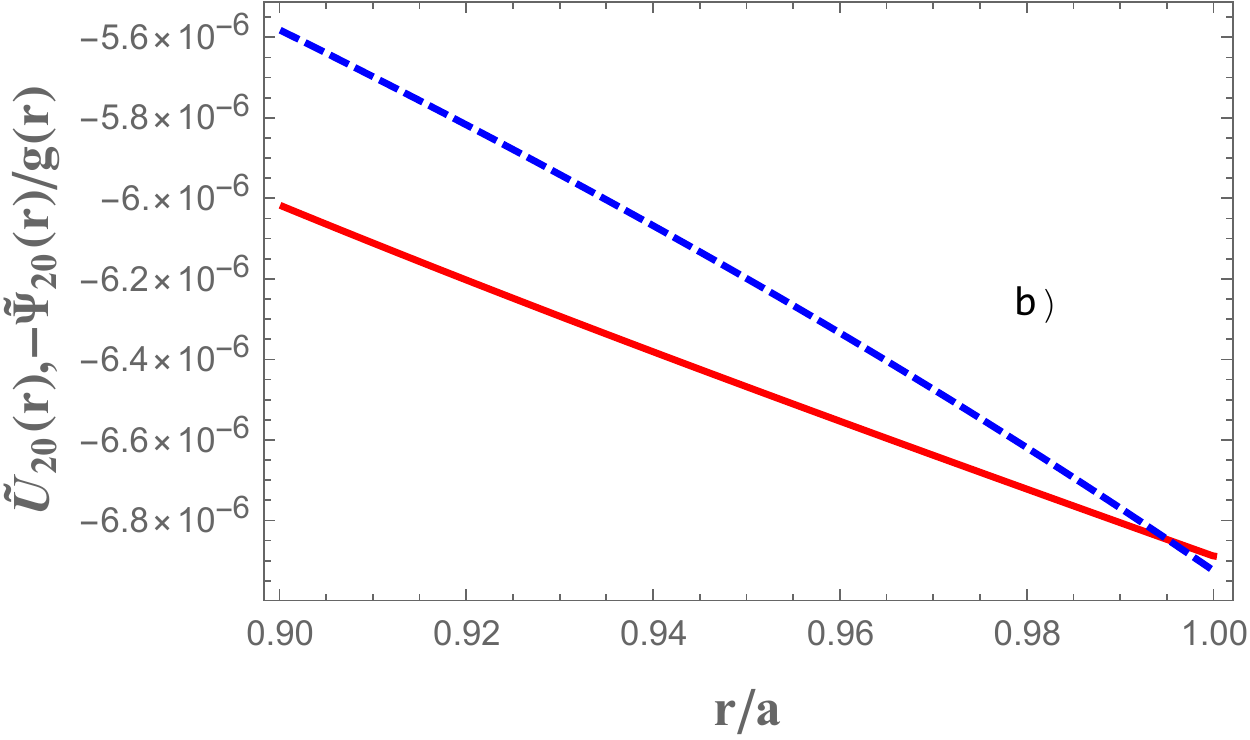}
        \includegraphics[width=\columnwidth]{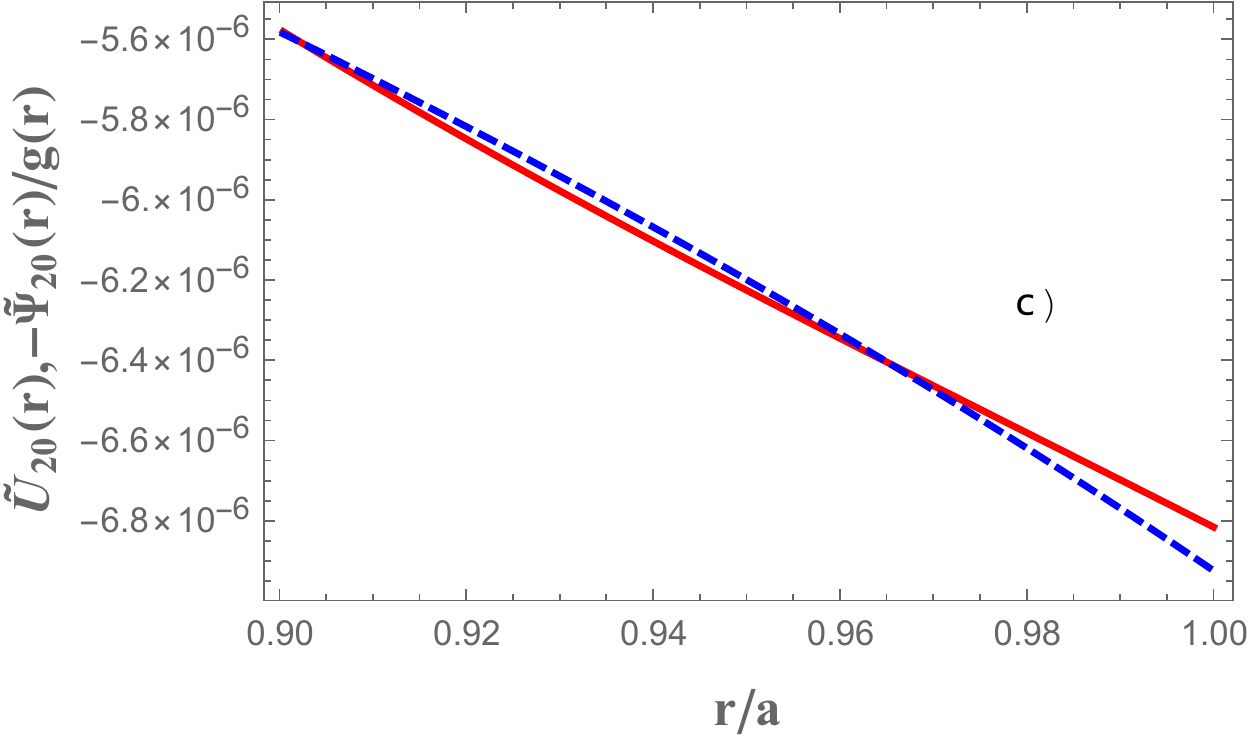}
        \includegraphics[width=\columnwidth]{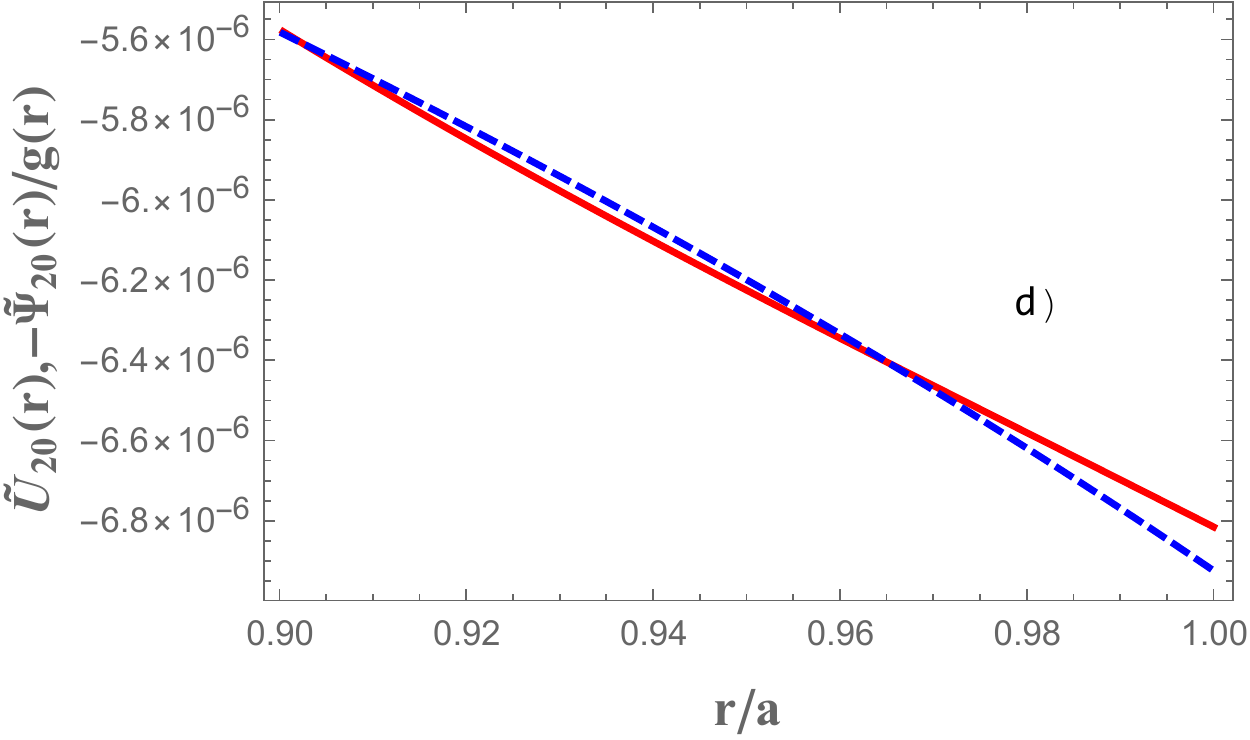}
\caption{Radial $\tilde{U}_{20}$ (red, solid) and geoid (blue, dashed) normalized displacements as a function of the normalized radius, from $r=r_c$ to $r=a$ and different values of the adiabatic index: $\gamma_f=\gamma_{eq}$ (a), $\gamma_f=2.1$ (b), $\gamma_f=200$ (c) and $\gamma_f=\infty$ (d).  
The curves for $\gamma_f=\infty$ and $\gamma_f=200$ are indistingishable. 
As the adiabatic index value goes to the incompressible limit, the radial displacement tends towards the geoid one.}
    \label{fig:strain vari gamma a r=rc}
\end{figure}

 {This change of behaviour of $\tilde{U}_{00}$ when going from the equilibrium limit to the incompressible one  can be explained by recalling that in a neutron star  \citep{chamel_livingreview}
\begin{equation}
\frac{\mu}{\kappa} \ll 1
\, .
\label{MU SU KAPPA}
\end{equation}
Therefore, the key physical aspects of the problem are already present by studying the $\mu\rightarrow0$ limit:  if the effective adiabatic index is such that $\gamma\neq\gamma_{eq}$ the Adams-Williamson equation \ref{wa generale} requires $\chi_{\ell m}$ to be zero, implying that there can be no volume changes. From Eq \eqref{frizzo} it follows that
\begin{equation}
R_{\ell m}=\kappa\chi_{\ell m}=0 
\end{equation}
and via Eq \eqref{FLUIDO TANGENZIALE} it is possible to show that the radial displacement must coincide with the geoid displacement defined in Eq \eqref{GEOID DISPLACEMENT}. 
Therefore,  we expect that 
\begin{equation}
    U_{\ell m}=-\frac{\Phi_{\ell m}}{g} = U^{geoid}_{\ell m} \, 
\end{equation}
when $\gamma\neq\gamma_{eq}$.
}
To check numerically this assertion, we first focus on the $\ell=2$ harmonic contribution  $U_{20}$ to the radial displacement and on the geoid radial displacement $U^{geoid}_{20}$.  
 {For the shear modulus in Eq \eqref{ANDAMENTO MU},  the difference between the radial and the geoid displacements decreases by increasing the adiabatic index, as can be seen in Fig \ref{fig:strain vari gamma a r=rc}: in the $\gamma_f=\gamma_{eq}$ case (panel-a) we have a clear departure from the geoid, but if $\gamma_f=200$ (panel-d) the two radial displacements almost coincide. As expected, the cases $\gamma_f=200$ and $\gamma_f=\infty$ give in practice the same result. 
}

\begin{figure}
		\includegraphics[width=\columnwidth]{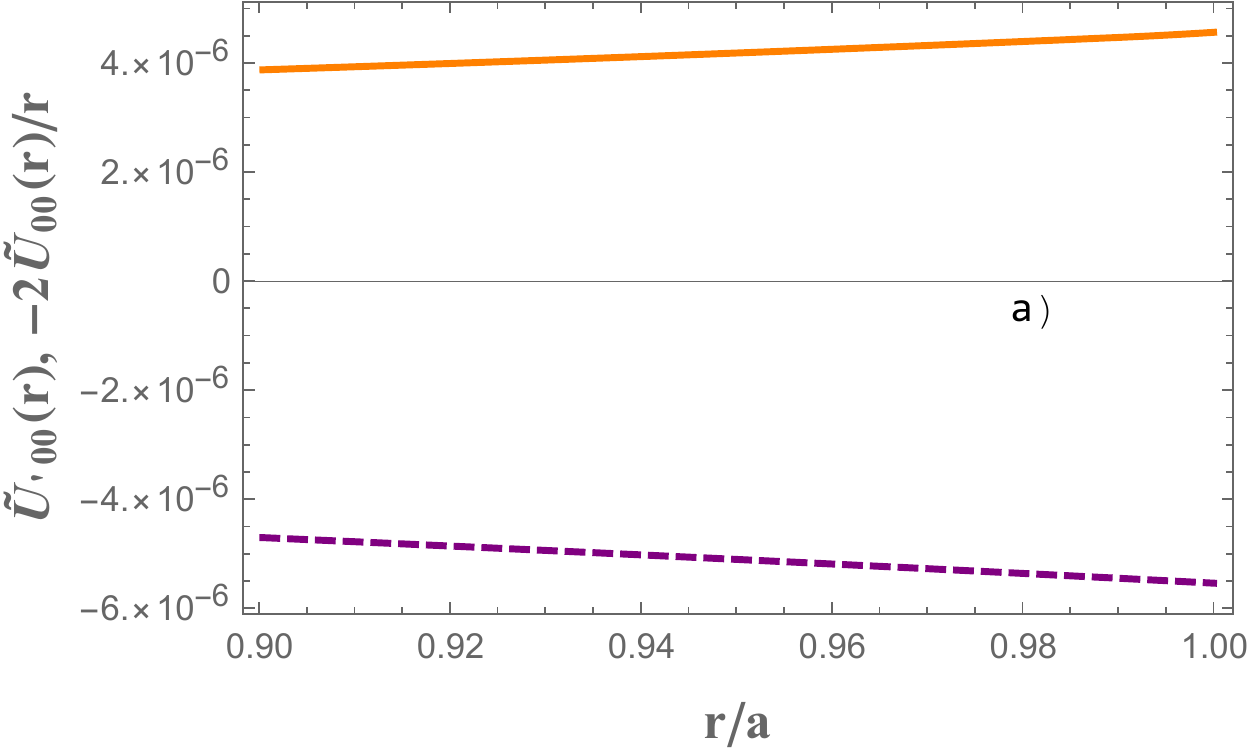}
        \includegraphics[width=\columnwidth]{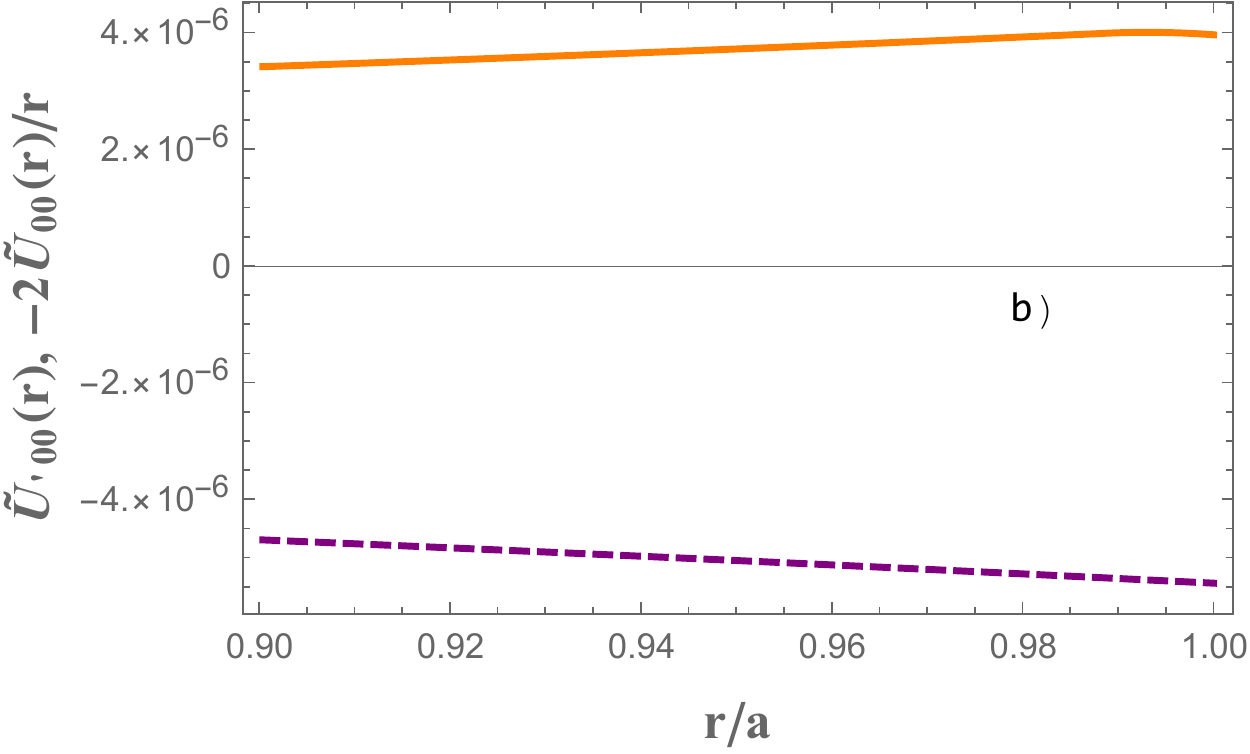}
        \includegraphics[width=\columnwidth]{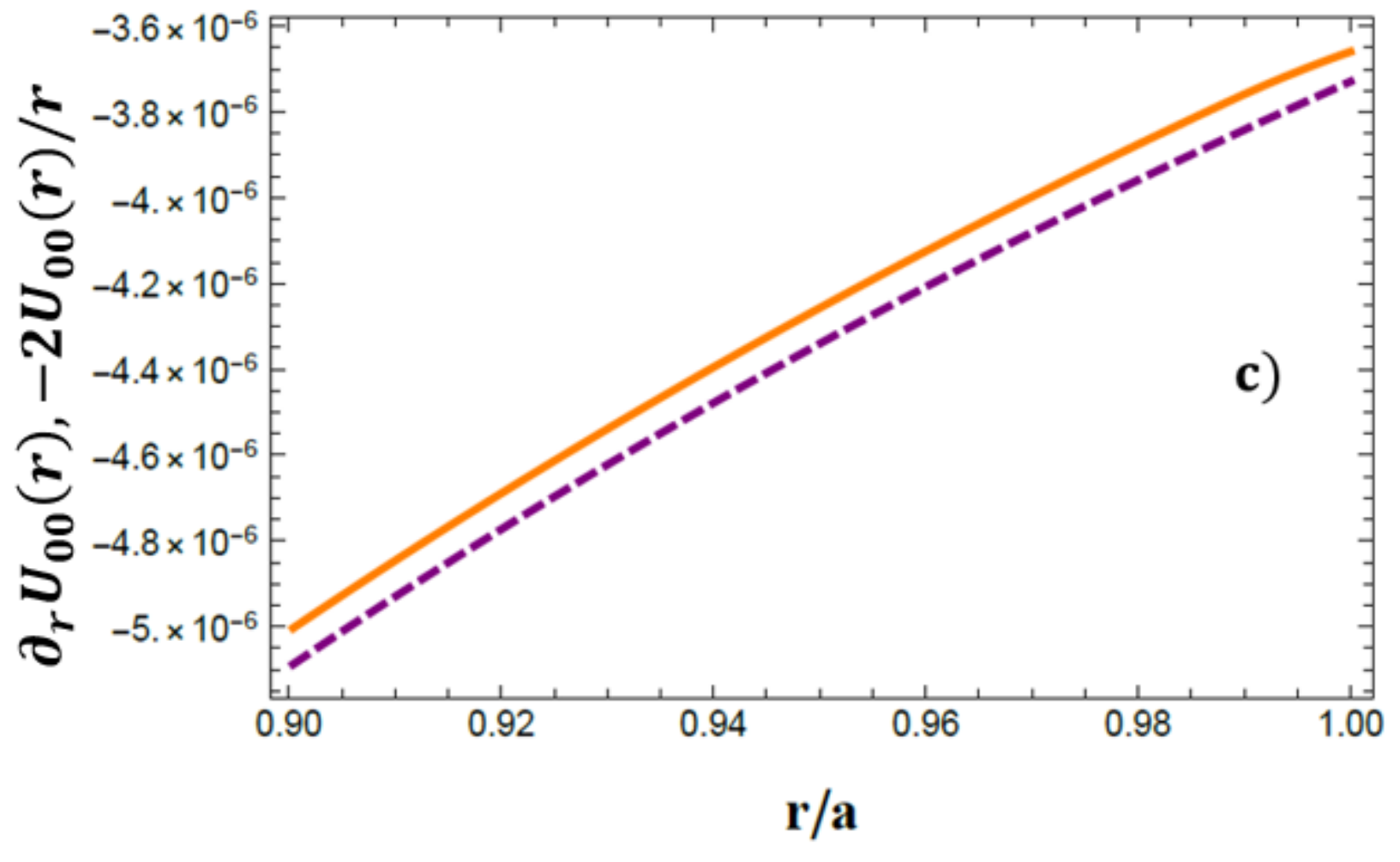}
        \includegraphics[width=\columnwidth]{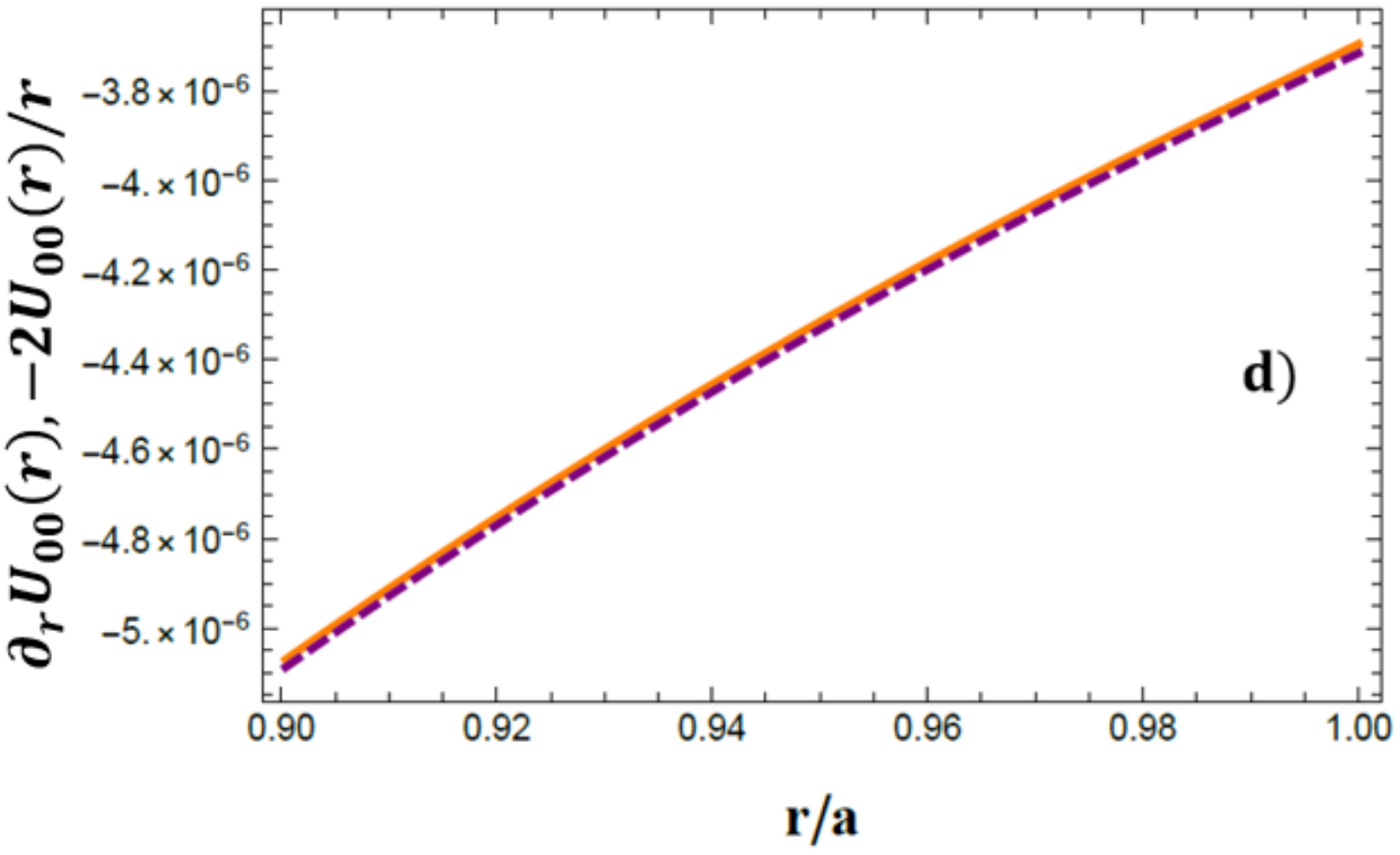}
\caption{ The functions $\partial_{r}U_{00}$ (orange) and $-2U_{00}/r$ (purple) calculated for  $\gamma_f=2$ (a), $\gamma_f=2.1$ (b), $\gamma_f=200$ (c) and $\gamma_f=\infty$ (d). As the adiabatic index increases the difference between $\partial_{r}U_{00}$ and $-2U_{00}/r$ goes to zero, as expected.}
    \label{fig:Uprimo meno u su r}
\end{figure}

Secondly, we numerically check the harmonic contribution of degree $\ell=0$  for increasing values of $\gamma_f$. 
 {From the analytical point of view, the incompressibility assumption  requires $\chi_{\ell m}=0$, implying that
\begin{equation}
\chi_{\ell m}=\partial_{r}U_{\ell m}+\frac{2}{r}U_{\ell m}-\frac{\ell\left(\ell+1\right)}{r}V_{\ell m}
=0\, .
\label{def Chi}
\end{equation}
Since the $\ell=0$ displacement is purely radial (i.e. $V_{\ell m}=0$), the above equation becomes
\begin{equation}
\partial_{r}U_{00}=-2\frac{U_{00}}{r}
\, .
\label{U0 primo}
\end{equation}
In  Fig \eqref{fig:Uprimo meno u su r} both $\partial_{r}\tilde{U}_{00}$ and $-2\tilde{U}_{00}/r$ are shown for the four different adiabatic indexes considered: again, we approach the incompressible limit  as $\gamma_f$  increases. }

 {
We can conclude that our numerical solution behaves as expected when varying the adiabatic index: by solving numerically the model we find that all the harmonic contributions to the radial displacement tend towards the geoid ones in the incompressible limit, as happens exactly in a purely fluid medium.}


\begin{figure}
		\includegraphics[width=\columnwidth]{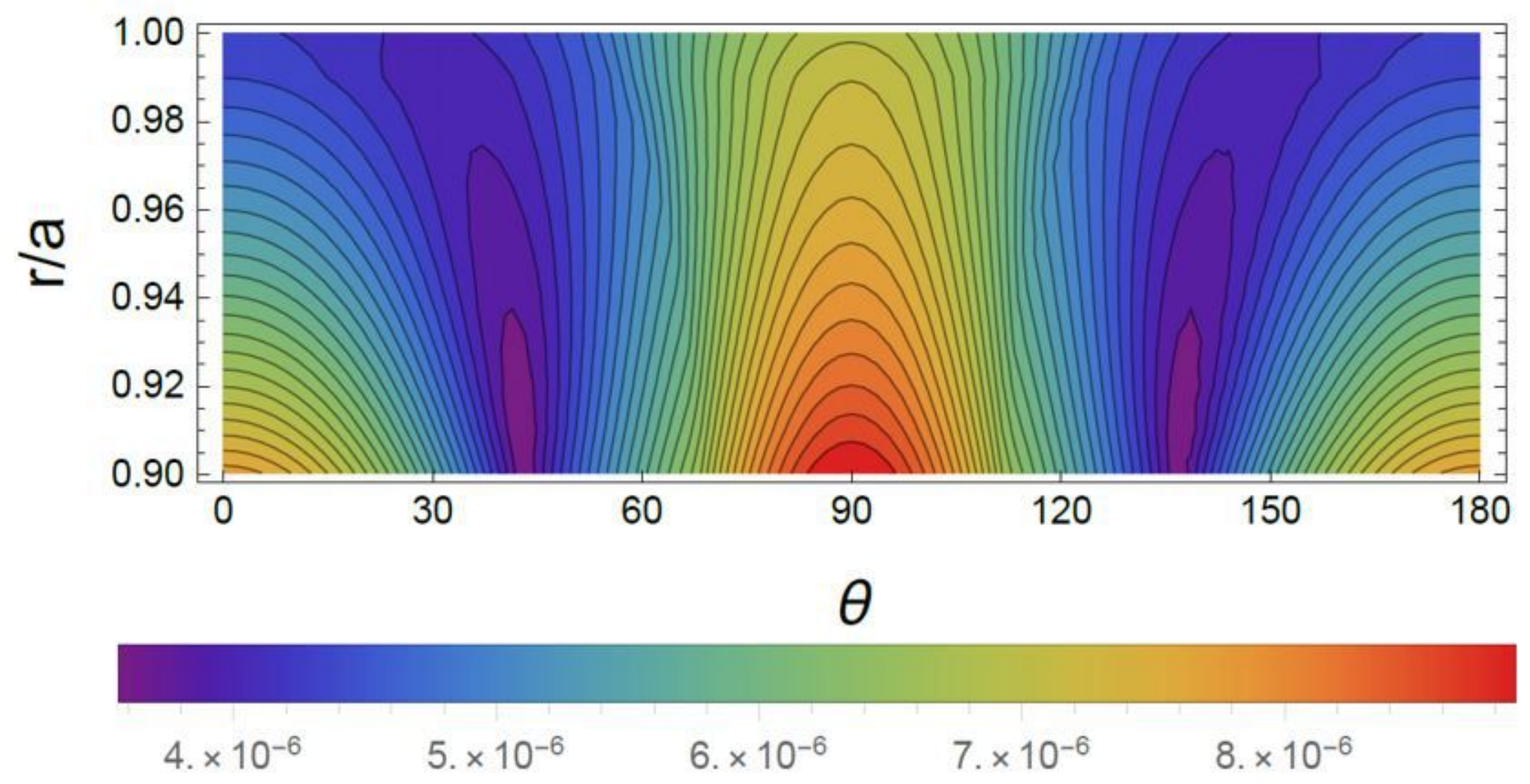}
        \includegraphics[width=\columnwidth]{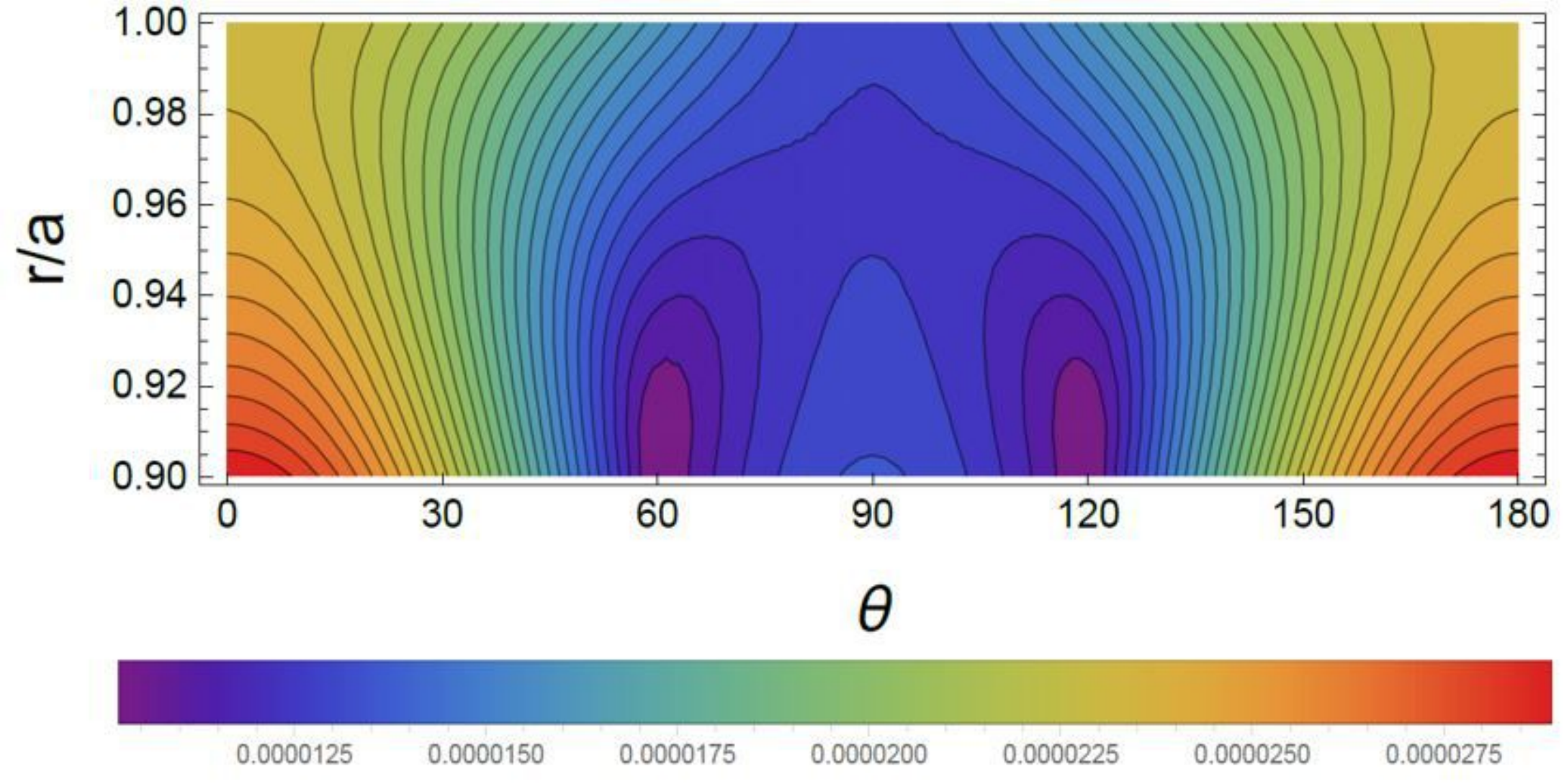}
    \caption{
    	Color map of the normalized strain angle $\tilde{\alpha}$ as a function of the colatitude $\theta$ and of the normalized radius $r/a$. The region shown here refers to the crustal layer, from $r=r_c$ to $r=a$. Our reference star of $M=1.4 \, M_{\odot}$ has been used, with different adiabatic indexes governing the perturbations: 
    	$\gamma=2.1$ (top) and $\gamma=200$ (bottom). Here $\alpha$ for $\gamma=\infty$ is not reported because it has the same shape and values of the case $\gamma=200$.
    	}
    \label{fig:Contour strain 1.4 gamma diversi}
\end{figure}

We now discuss the strain angle in the crust. In Fig \ref{fig:Contour strain 1.4 gamma diversi} we plot the normalized strain angle $\tilde{\alpha}$ as a function of $r$ and $\theta$. 
The color  maps representing the various values of $\tilde{\alpha}$ differ significantly when going from the equilibrium to the non-equilibrium configurations.

The value of $\gamma_f$ influences the slope of the strain angle curve: if for $\gamma_{f}=2$, $\alpha$ is an increasing function of $r$  {(see Fig \ref{fig:Contour strain 1.4 gamma 2})}, for $\gamma_{f}=2.1$, $200$ and $\infty$ the opposite is true.   {This is an interesting result, since one could expect the strain to be maximum at the star's surface, where $\mu$ is smaller \citep{cutler2003}. However, the shear modulus is so small with respect to the bulk modulus that the global crustal deformation is ruled essentially by the value of the adiabatic index, so that even for small deviations of $\gamma_f$ from the equilibrium value the strain angle behaves qualitatively as in the incompressible model studied in \cite{giliberti2019PASA}.}

\begin{figure}
        \includegraphics[width=\columnwidth]{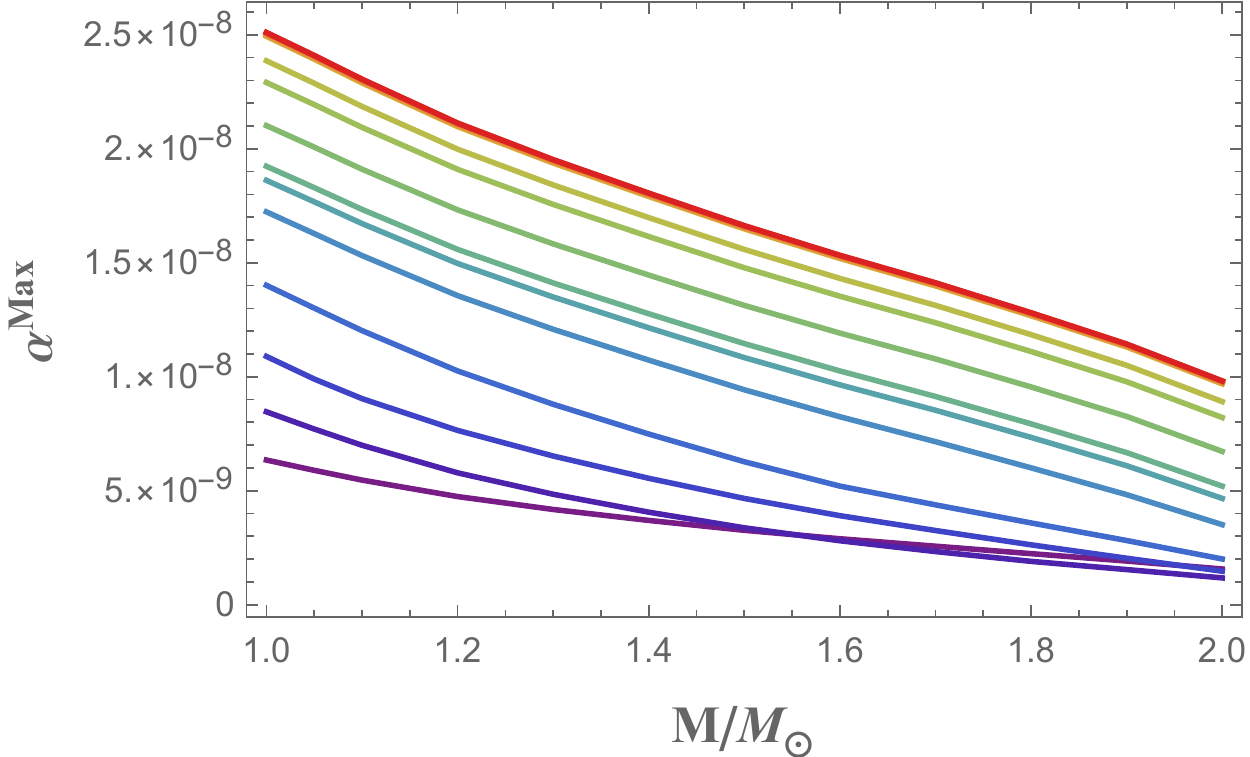}
\caption{
	Maximum strain angle $\alpha^{Max}$ as a function of the stellar mass, calculated for $\Omega=1\,\mathrm{rad/s}$ and different adiabatic indexes ($\gamma_{f}=$2, 2.05, 2.1, 2.2, 2.5, 2.8, 3, 4, 7, 12, 100 and 1000) going monotonically from the lowest purple line ($\gamma_{f}=2$) to the red one on top ($\gamma_{f}=1000$).  The curve for $\gamma_{f}\geq 100$ is superimposed to the one for $\gamma_{f}\geq 1000$, indicating that  the star's behaviour is essentially indistinguishable with respect to the analytical incompressible limit. The lowest curve (purple) coincides with the curve in Fig \ref{fig:Strain vari masse gamma equilibrio}.
	}
    \label{fig:strain vari gamma varie masse}
\end{figure}

Finally, we estimate the importance of the stellar mass parameter $M$ by studying the function $\alpha^{Max}(M)$. 
 {As in section 5.1 we have to evaluate the coefficient $d$ for stellar configurations of different mass and we fix $\Omega=1 \,\mathrm{rad/s}$, so that the results can easily rescaled for different rotation rates.} 
We find that the strain angle dependence on mass is almost the same for every value of the frozen adiabatic index. We checked this behaviour by employing  a larger set of values for $\gamma_f$, with values ranging from $\gamma_{f}=\gamma_{eq}$ to the incompressible limit, as reported in Fig  \ref{fig:strain vari gamma varie masse}.

\begin{figure}
        \includegraphics[width=\columnwidth]{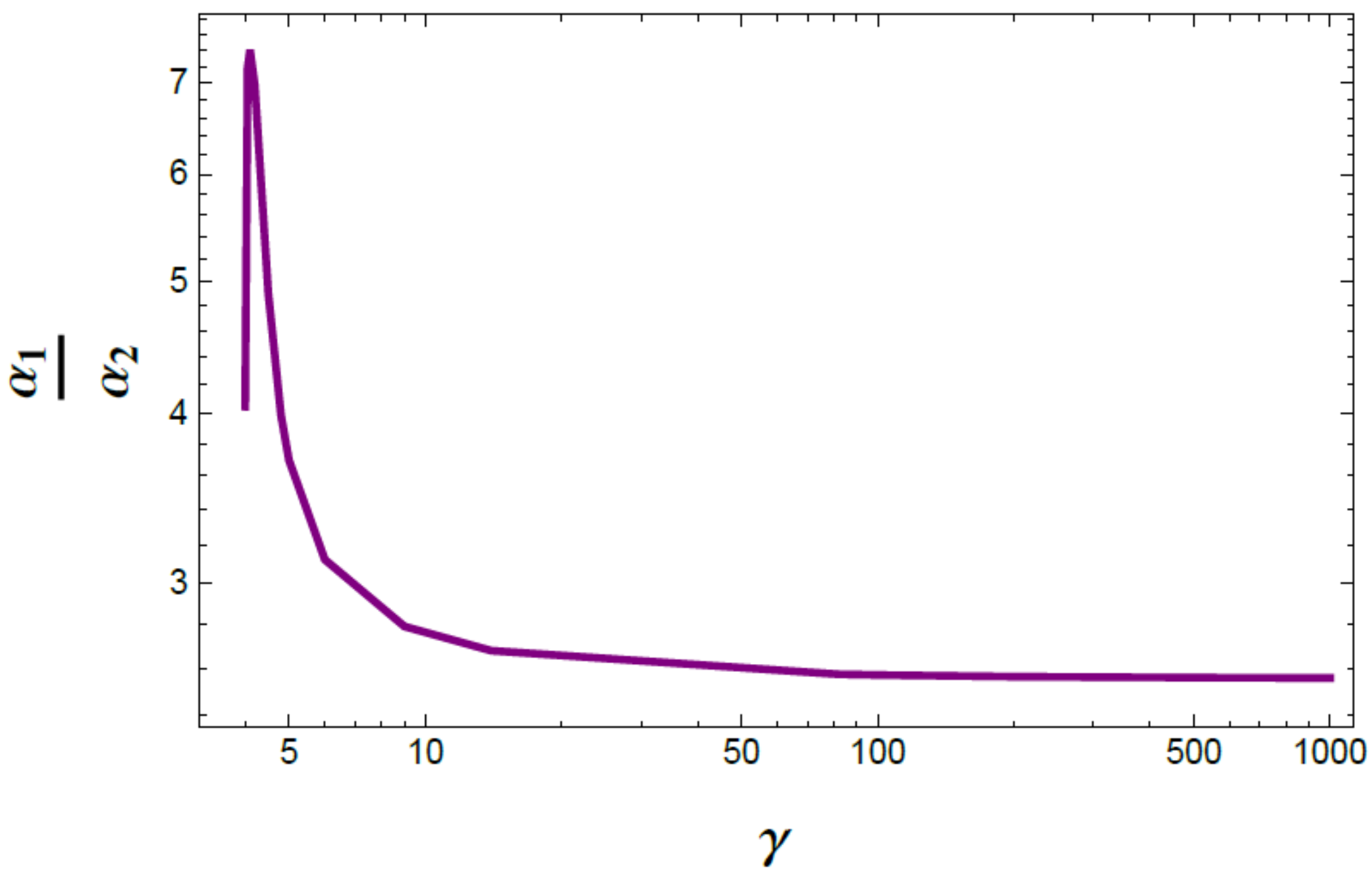}
  \caption{Ratio of the maximum strain angles for a low-mass and high-mass NS, $\alpha_{M=1M_{\odot}}^{Max}/\alpha_{M=2M_{\odot}}^{Max}$, as a function of the adiabatic index $\gamma_f$ for $\Omega=1$ rad/s. }
    \label{fig:ratio a1/a2}
\end{figure}

The comparison between different adiabatic indexes allows to calculate the ratio $\alpha^{Max}(1 M_\odot)/\alpha^{Max}(2 M_\odot)$ for different scenarios, going from the equilibrium to the incompressible one. This is shown in Fig \ref{fig:ratio a1/a2}. 
As we can see that ratio has large values when a $\gamma_f$ near $\gamma_{eq}$ is employed, but it rapidly decreases towards the asymptote $\alpha^{Max}(1 M_\odot)/\alpha^{Max}(2 M_\odot)\simeq2.6$ as the incompressible limit is approached. 
This latter value resembles the one obtained with the homogeneous two-density incompressible model where $\alpha^{Max}(1 M_\odot) / \alpha^{Max}(2 M_\odot) \simeq 2\div3$  \citep{giliberti2019PASA}.

\subsection{Effective adiabatic index: the effect of General Relativity}

We end this section with a technical note, expanding the motivation behind our basic working assumption of a non-relativistic framework.

The Adams-Williamson equation \eqref{ADAMS-WILLIAMSON} tells us that the adiabatic index (both the equilibrium one as well as the more uncertain frozen one) depends strongly on the stratification: this can be envisaged by comparing the \emph{effective} adiabatic index of Eq \eqref{gamma effettivo}, calculated by using the equilibrium configuration of a non-rotating star with the equilibrium one that is intrinsic to the barotropic EoS used.

\begin{figure}
        \includegraphics[width=\columnwidth]{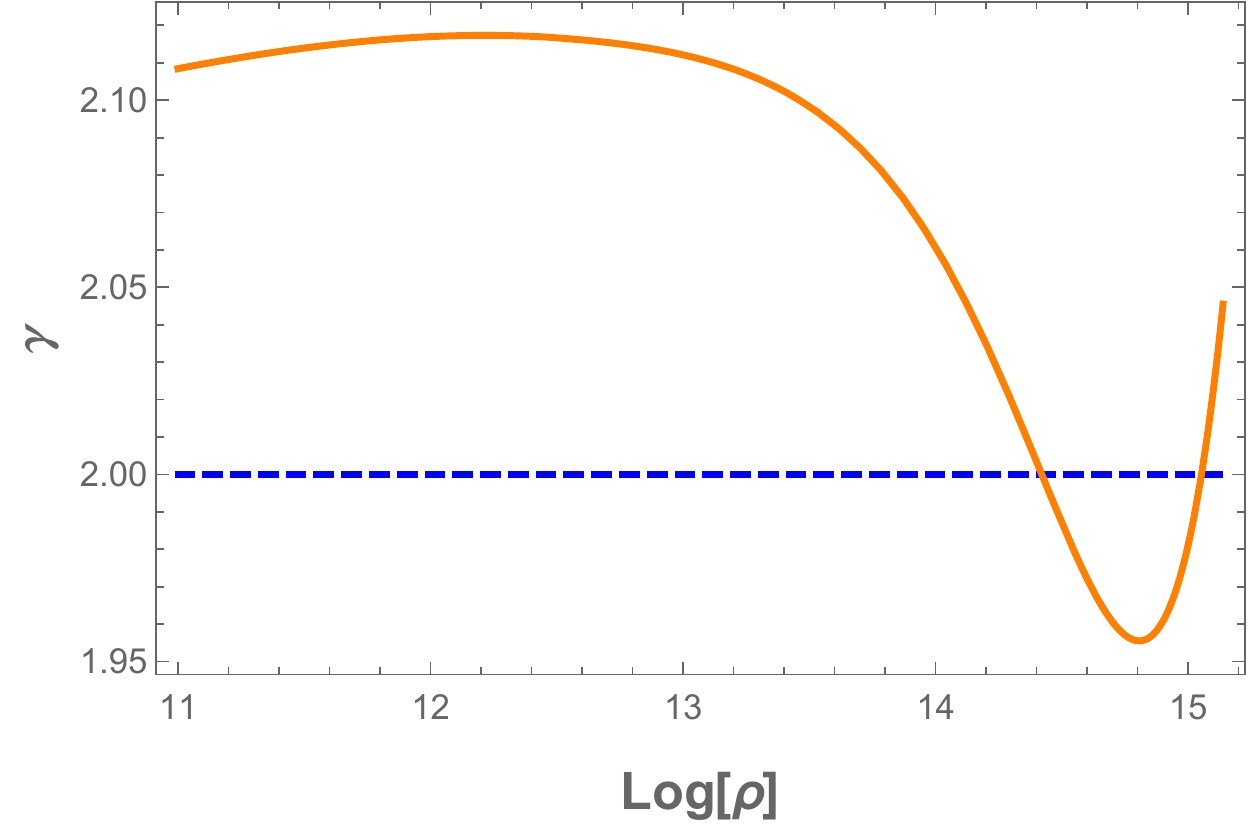}
  \caption{
  	The effective adiabatic index calculated from the equilibrium configuration in General Relativity (orange) and in  Newtonian gravity (blue, dashed). 
  	Both the effective adiabatic indexes are shown for a  $M=1.4M_{\odot}$ neutron star which EoS is given by the polytrope $n=1$.
  	}
    \label{fig:gamma New vs GR}
\end{figure}

We consider the usual  $M=1.4 \, M_{\odot}$ star described by a polytrope $n=1$, which stratification has been calculated both assuming Newtonian gravity and using the TOV equations. The difference between the two effective adiabatic indexes is shown in Fig \ref{fig:gamma New vs GR}.

In the Newtonian case the effective index is clearly given by $\gamma=\gamma_{eq}$, as it should be. However, when the radial profiles are integrated by using the TOV equations, there is a departure of the effective adiabatic index $\gamma$ from the equilibrium value, exceeding the $5\%$ in the crust. 

As  discussed in the previous section, the behavior of the stresses and displacements calculated for $\gamma_f=2.1$ is different from the equilibrium case. 
 {This warns against using an equilibrium stellar configuration calculated via the TOV equations in a Newtonian model for stresses and strains: in this way an artificial effective adiabatic index that does not correspond to the physical response of the star is introduced.}

\section{Application to pulsar glitches}

Glitches are sudden jumps in the rotational frequency of a pulsar followed by a period of slow recovery that can last for several weeks or months.
 {
	The first attempts to explain the glitches activity indicated starquakes as a possible mechanism, but soon it became clear that large glitches, as the one of the Vela pulsar, cannot be only due to a change of the NS moment of inertia.} \citep{baym1969}. 
 { Nowadays it is thought that glitches are due to an angular momentum exchange between the superfluid core and the elastic crust, following  a sudden unpinning of many  vortices contained in the superfluid permeating the inner crust and the outer core of a NS} \citep{haskell_rev_glitches}. 
 {Within this latter scenario, the failure of the crust can still play a substantial role as a possible mechanism that triggers the unpinning of superfluid vortices. We can employ our model to investigate this possibility by studying the stresses that can be accumulated during the spin-down of a glitching pulsar.}

Following the same approach of \cite{giliberti2019PASA}, we use the present and more refined model to estimate the accumulated strain due to the spin-down between two  glitches in the Vela, which parameters  $\Omega$ and spin-down rate $\dot \Omega$ are given in Tab \ref{tab Vela}.

\begin{table}
	\centering{}%
	\begin{tabular}{|c|c|c|}
		\hline
		$\Omega$ (rad/s) & $\dot{\Omega}$ (rad/s$^{2}$) & $\omega$ (rad/s)\tabularnewline
		\hline 
		70.338 & -9.846$\times 10^{-11}$ & 80.44$\times 10^{-4}$\tabularnewline
	\end{tabular}\caption{
	The rotational  parameters and the average observed waiting time between two large glitches in the Vela pulsar.
}
\label{tab Vela}
\end{table}

The absolute difference in angular velocity between two  Vela glitches can be estimated as $\omega \approx |\dot{\Omega}|  \langle t_{gl}\rangle $, where $\langle t_{gl}\rangle\approx 3\,$yr is the typical observed inter-glitch time. 
 { Assuming that the only loading mechanism is due to the slowly-changing angular velocity, the strain developed during the inter-glitch time is roughly given by} 
\begin{equation}
\alpha
=    \,
(d\left(\Omega\right)-d\left(\Omega-\omega\right)) 
\, \tilde{\alpha} \,
\approx   \, \frac{2 \Omega \omega a^2}{3 \, v^2} \, \tilde{\alpha}
\,  ,
\end{equation}
where $\tilde{\alpha}$ is the normalized strain angle and $d(\Omega)$ is given in Eq \eqref{d}.
Using the fiducial parameters in Table \ref{tab Vela} and the rotational parameters in Table 2, we get
\begin{equation}
\alpha_{Vela}^{Max}=3.5\times10^{-4}\tilde{\alpha}
\, ,
\end{equation}
which means that the strain accumulated due only to the spin-down between two glitches is of the order of 
\begin{equation}
\alpha_{Vela}^{Max} \, = \, 2.1\times10^{-9} \, .
\end{equation}
This is an  extremely small value if compared to the assumed breaking strain in the range $10^{-5}\div10^{-1}$: therefore, the crust's failure may be a viable trigger for glitches only in the eventuality that the crust is always stressed and very near to the breaking threshold. In other words,  in order to trigger a sequence of glitches, the crust-quakes must release only an extremely small portion of the crustal stresses that have been accumulated up to that point.

Alternatively, very intense internal loads other than the pure spin-down process should take part in the process, like the pinning of many vortex lines in the crust \citep{rudermanI1991}, a possibility that will be investigated in a future work.   

\section{Conclusion}

The first aim of the present work is to introduce a Newtonian model to study the deformation of a self-gravitating and compressible neutron star under a variety of loads, like tidal forces and non-conservative forces. In this sense, we considerably extended the model of \cite{giliberti2019PASA}, in which the NS has been modeled as an incompressible body with two uniform layers. On the other hand, in the present treatment we considered a continuous stratification of matter that is consistent with the assumed EoS and self-gravitation.

 {
	In the numerical analysis of the model,  the equilibrium structure of the NS has been fixed by assuming a polytropic EoS with $n=1$ (i.e. $\gamma_{eq}=2$), which mass-radius relation has been set to be the one of the SLy EoS \citet{douchin2001}. 
In this way our Newtonian model is consistent, in the sense that no spurious effects due to an artificial effective adiabatic index are introduced (as explained in section 5.3), but the mass and radius of the star have realistic values obtained via the integration of the TOV equations.
In particular, we studied the dependence of the maximum strain angle $\alpha^{Max}$ on the stellar mass for compressible NS, finding that the strain caused by the change in the rotation rate of a pulsar is a decreasing function of the stellar mass. Hence, the crust of a very compact and massive NS is  more difficult to deform and, ultimately, to break with respect to a $\sim 1 M_\odot$ NS.}

As a first application of our model, we studied the effect of a possible departure of the adiabatic index from its equilibrium value \citep{HL2002} during the build-up of stresses induced by the slowly changing rotation rate of the NS.
The elastic response of a neutron star in the case of a non-equilibrium adiabatic index $\gamma_f$ shows interesting features. We find that, despite the small difference in the value of adiabatic indexes in the cases $\gamma_{eq}=2$ and $\gamma_{f}=2.1$, the dynamical responses of the model are clearly distinguishable, either for displacement,  stresses and strain angles. 
This fact is closely related to the smallness of the shear modulus with respect to the bulk modulus in a NS, as discussed in section 5.2. 


Secondly, we checked that the sensitivity of strains and stresses to the actual value of the adiabatic index governing perturbations puts a severe warning against the naive use of density profiles obtained via integration of the TOV equations in Newtonian models for stellar deformations. 
In fact, given the same EoS, the effective adiabatic index calculated by employing the relativistic stratification turns out to be quite different with respect to the value $\gamma_{eq}$, that is intrinsic to the EoS.
Therefore, in order to clearly separate the effect of inconsistent stratification from to the one arising from the use of an frozen adiabatic index, a completely relativistic model is needed. 

 {
Finally, we explored to what extent the spin-down of a pulsar between two consecutive glitches can load the solid crust. With the present, much more refined model for stellar deformations, we confirm the results of \cite{giliberti2019PASA}: the loading of the solid crust caused by the spin-down between glitches is not enough to reach the breaking threshold and trigger a crustquake.} In fact, the difference of angular velocity  between glitches in the Vela gives rise to extremely small maximum strain angles, of the order of $\alpha^{Max} \sim \left(\Omega\omega/1\,\mathrm{rad^{2}s^{-2}}\right)\times10^{-9}$ for a typical $M=1.4 \, M_{\odot}$ NS. 
These values are orders of magnitude smaller even with respect to the smallest breaking strain that is theoretically expected, of the order of $10^{-5}$. 

 {This result challenges the idea that crustquake are the cause of the sudden unpinning of the superfluid vortices in the crust and, ultimately, that crustquakes are the trigger of vortex-mediated glitches in pulsars: this could be a possibility only if the crust never relieves all the accumulated stresses (i.e. the strain angle in some regions of the crust is always close to the breaking threshold) or some  internal load that is  orders of magnitude stronger than the variation in the centrifugal force (maybe due to vortex pinning or to the magnetic field) is operating in the crust.}



\section*{Acknowledgements}

We acknowledge the PHAROS COST Action (CA16214) and INFN for partial support. MA acknowledges support from the Polish National Science Centre grant SONATA BIS 2015/18/E/ST9/00577, P.I.: B. Haskell. 
We thank Morgane Fortin, Leszek Zdunik and Pawe{\l }  Haensel for valuable discussion. 	
	
	
\bibliographystyle{mnras}
\bibliography{biblio_articolomodi}


\appendix
\section{\lowercase{spherical harmonics}}
In this work spherical harmonics are defined as
\begin{equation}
Y_{\ell m}\left(\theta,\phi\right)=P_{\ell m}\left(\cos\theta\right)e^{im\phi},
\end{equation}
where $P_{\ell m}$ are the Legendre polynomials, given by 
\begin{align*}
&P_{\ell m}(x)=\frac{(1-x^{2})^{m/2}}{2^{\ell}\ell!}\frac{d^{\ell+m}\left(x^{2}-1\right)^{\ell}}{dx^{\ell+m}}& \quad&\text{if}\quad  m\geq 0
\\
&P_{\ell m}(x)=\left(-1\right)^{m}\frac{\left(\ell-m\right)!}{\left(\ell+m\right)!}P_{\ell m}\left(x\right)& \quad &\text{if}\quad  m< 0 \, .
\end{align*}
The spherical harmonics are the eigenvalues of the angular part of the Laplacian, i.e.
\begin{equation}
\nabla^{2}Y_{\ell m}=-\frac{\ell\left(\ell+1\right)}{r^{2}}Y_{\ell m},
\end{equation}
and are normalized as
\begin{equation}
\intop_{\Omega} Y_{\ell m} Y_{\ell m}^{*} d\Omega = \frac{4\pi}{2\ell+1} \frac{\left(\ell+m\right)!}{\left(\ell-m\right)!} \delta_{\ell \ell^{'}} \delta_{m m'} \, ,
\end{equation}
where $ d\Omega=\sin\theta d\theta d\phi$.
Consistently with the previous definitions, we expand the total incremental potential $\Phi^{\Delta}$ (as well as all the scalar functions) as
\begin{equation}
\Phi^{\Delta}\left(r,\theta,\phi\right)=\sum_{\ell=0}^{\infty}\sum_{m=-\ell}^{\ell}\Phi_{\ell m}\left(r\right)Y_{\ell m}\left(\theta,\phi\right) \, .
\label{eq:philm}
\end{equation}
The expansion of vectorial quantities is more subtle. In particular, the total displacement $\mathbfit{u}$ is decomposed in terms of the spheroidal $\boldsymbol{u}_S$ and the toroidal $\boldsymbol{u}_T$ displacements as
\begin{equation}
\boldsymbol{u}=\boldsymbol{u}_{S}+\boldsymbol{u}_{T} \, ,
\label{eq:uTS}
\end{equation}
where
\begin{align}
& \boldsymbol{u}_{S}\left(r\right)=\sum_{\ell=0}^{\infty}\sum_{m=-\ell}^{\ell}\left[U_{\ell m}\left(r\right)\boldsymbol{R}_{\ell m}\left(\theta,\varphi\right)+V_{\ell m}\left(r\right)\boldsymbol{S}_{\ell m}\left(\theta,\varphi\right)\right]
\label{eq:disp_harminics1}
\\
& \boldsymbol{u}_{T}\left(r\right)=\sum_{\ell=0}^{\infty}\sum_{m=-\ell}^{\ell}\left[W_{\ell m}\left(r\right)\boldsymbol{T}_{\ell m}\left(\theta,\varphi\right)\right].
\label{eq:disp_harminics2}
\end{align}
In the above expansions, the symbols $\boldsymbol{R}_{\ell m}$, $\boldsymbol{S}_{\ell m}$, $\boldsymbol{T}_{\ell m}$ are vectorial quantities defined by
\begin{align}
& \boldsymbol{R}_{\ell m}=Y_{m}\boldsymbol{e}_{r}
\\
& \boldsymbol{S}_{\ell m}=r\boldsymbol{\nabla}Y_{\ell m}=\partial_{\theta}Y_{\ell m}\boldsymbol{e}_{\theta}+\frac{1}{\sin\theta}\partial_{\varphi}Y_{\ell m}\boldsymbol{e}_{\varphi}
\\
& \boldsymbol{T}_{\ell m}=\boldsymbol{\nabla}\times\left(\boldsymbol{r}Y_{\ell m}\right)=\frac{1}{\sin\theta}\partial_{\varphi}Y_{\ell m}\boldsymbol{e}_{\theta}-\partial_{\theta}Y_{\ell m}\boldsymbol{e}_{\varphi} \, ,
\end{align}
where $\boldsymbol{e}_{r},\boldsymbol{e}_{\theta}$ and $\boldsymbol{e}_{\varphi}$ are the usual unit vectors of the spherical coordinate system.

The incremental stress acting on a spherical surface element with outward normal $\boldsymbol{e}_{r}$ can be computed as
\begin{equation}
\boldsymbol{\sigma}^{\delta}\cdot\boldsymbol{e}_{r}=\sum_{\ell m}\left(R_{\ell m}\boldsymbol{R}_{\ell m}+S_{\ell m}\boldsymbol{S}_{\ell m}+T_{\ell m}\boldsymbol{T}_{\ell m}\right),
\end{equation}
where 
\begin{align}
& R_{\ell m}=\lambda\chi_{\ell m}+2\mu\partial_{r}U_{\ell m} \, ,
\\
& S_{\ell m}=\mu\left(\partial_{r}W_{\ell m}+\frac{U_{\ell m}-V_{\ell m}}{r}\right) \, ,
\\
& T_{\ell m}=\mu\left(\partial_{r}W_{\ell m}-\frac{W_{\ell m}}{r}\right) \, .
\end{align}
In the main text, $R_{\ell m}$ and $S_{\ell m}$ are referred to as the \emph{radial} and  \emph{tangential spheroidal} stresses respectively. 
On the other hand, $T_{\ell m}$ is the \emph{toroidal stress}. 

Finally, a generic non-conservative force $\boldsymbol{h}$ can be expanded in terms of three real and independent sets of coefficients $h_{\ell m}^{R}$, $h_{\ell m}^{S}$ and  $h_{\ell m}^{T}$ according to the formula
\begin{equation}
\boldsymbol{h}
=
\sum_{\ell=0}^{\infty} \sum_{m=-\ell}^{\ell}
\left(
h_{\ell m}^{R} \boldsymbol{R}_{\ell m}+h_{\ell m}^{S} \boldsymbol{S}_{\ell m}
+h_{\ell m}^{T} \boldsymbol{T}_{\ell m}
\right) \, .
\label{eq:A13}
\end{equation}
\section{\lowercase{General form of the matrix} $A$}
In Eq \eqref{EQUAZIONE FONDAMENTALE EQUILIBRIO}, the spheroidal equations are written in a compact notation thanks to the use of the $6 \times 6$ matrix $A_{\ell}\left(r\right)$, which general form is 
\[
A_{\ell}\left(r\right)=\left(\begin{array}{ccc}
-\frac{2\lambda}{r\beta} & \frac{\ell\left(\ell+1\right)\lambda}{r\beta} & \frac{1}{\beta}\\
-\frac{1}{r} & \frac{1}{r} & 0\\
\frac{4}{r}\left(\frac{3\kappa\mu}{r\beta}-\rho_{0}g\right) & \frac{\ell\left(\ell+1\right)}{r}\left(\rho_{0}g-\frac{6\kappa\mu}{r\beta}\right) & -\frac{4}{r}\frac{\mu}{\beta}\\
\frac{1}{r}\left(\rho_{0}g-\frac{6}{r}\frac{\mu\kappa}{\beta}\right) & \frac{2\mu}{r^{2}}\left[\ell\left(\ell+1\right)\left(1+\frac{\lambda}{\beta}\right)-1\right] & -\frac{\lambda}{r\beta}\\
-4\pi G\rho_{0} & 0 & 0\\
-\frac{4\pi G\rho_{0}\left(\ell+1\right)}{r} & \frac{4\pi G\rho_{0}\ell\left(\ell+1\right)}{r} & 0
\end{array}\right.
\]
\begin{equation}
\left.\begin{array}{ccc}
0 & 0 & 0\\
\frac{1}{\mu} & 0 & 0\\
\frac{\ell\left(\ell+1\right)}{r} & -\frac{\rho_{0}\left(\ell+1\right)}{r} & \rho_{0}\\
-\frac{3}{r} & \frac{\rho_{0}}{r} & 0\\
0 & -\frac{\ell+1}{r} & 1\\
0 & 0 & \frac{\ell-1}{r}
\end{array}\right)
\end{equation}
where $\beta=\kappa+\frac{4}{3}\mu$ and $\lambda=\kappa-\frac{2}{3}\mu$.

\section{\lowercase{the $\ell=0$ harmonic}}

The matrix $A_{\ell}\left(r\right)$, for the harmonic $\ell=0$ is
\begin{equation}
A_{0}\left(r\right)=\left(\begin{array}{cccc}
-\frac{2\left(\kappa-\frac{2}{3}\mu\right)}{r\beta} & \frac{1}{\beta} & 0 & 0\\
\frac{4}{r}\left(\frac{3\kappa\mu}{r\beta}-\rho_{0}g\right) & -\frac{4}{r}\frac{\mu}{\beta} & -\frac{\rho_{0}}{r} & \rho_{0}\\
-4\pi G\rho_{0} & 0 & -\frac{1}{r} & 1\\
-\frac{4\pi G\rho_{0}}{r} & 0 & 0 & -\frac{1}{r}
\end{array}\right).
\end{equation}
This expression can be obtained from the general form for the matrix $A$ (see appendix B) by putting $\ell=0$ and neglecting the tangential displacement and stress. The equation that we have to solve is
\begin{equation}
\frac{d\boldsymbol{w}_{00}}{dr}=\boldsymbol{A}_{0}\boldsymbol{w}_{00}+\boldsymbol{f}_{cen},
\label{EQUAZIONE L=0}
\end{equation}
where $\boldsymbol{f}_{cen}$ is the centrifugal force vector 
\begin{equation}
\boldsymbol{f}_{cen}=(0,f_{cen},0,0)
\end{equation} 
and the spherical solution $\boldsymbol{w}_{00}$ is the four-vector  
\begin{equation}
\boldsymbol{w}_{00}\left(r\right)
=
\left(U_{00}\left(r\right),R_{00}\left(r\right),\Phi_{00}\left(r\right),Q_{00}\left(r\right)\right).
\end{equation}
We underline that in this case the approach is different with respect to the one presented in Eq \eqref{TERMINI POTENZIALE DELTA} for $\ell\geq2$, since here the potential $\Phi^{\Delta}$ coincides with the perturbed gravitational potential $\phi^{\Delta}$. 
It can be shown that the system \eqref{EQUAZIONE L=0} can be simplified in a system for $U$ and $R$. In this respect the general vector solution of Eq \ref{EQUAZIONE L=0} can be written as
\begin{equation}
\boldsymbol{w}_{00}\left(r\right)=\left(\begin{array}{cc}
U^{reg}\left(r\right) & 0\\
R^{reg}\left(r\right) & 0\\
\Phi^{reg}\left(r\right) & 1\\
\frac{\Phi^{reg}}{r}\left(r\right) & \frac{1}{r}
\end{array}\right)\left(\begin{array}{c}
C_{1}\\
C_{2}
\end{array}\right)+\left(\begin{array}{c}
U^{f}\\
R^{f}\\
\Phi^{f}\\
Q^{f}
\end{array}\right) \, .
\label{SOLUZIONE GENERALE}
\end{equation}
The superscript $reg$ indicates the regular solution of the associated homogeneous system; while $f$ indicates a particular solution of \eqref{EQUAZIONE L=0}. Note that the constant $C_{2}$ does not affect the radial displacement and the radial stress, that can be found by fixing the other constant $C_{1}$. This can be done by imposing the boundary conditions for the radial stress, namely
\begin{equation}
R_{00}\left(a\right)=0.
\end{equation}
To find the solutions of Eq \eqref{SOLUZIONE GENERALE}, we need some initial conditions. Here, for simplicity, we are looking only for the explicit solution for $U_{00}$ and $R_{00}$. In the innermost part of the star, we can state that all the quantities $\rho$, $\kappa$ and $\mu$ are constant, therefore
\begin{equation}
\rho=\rho_{0}, \quad \kappa=\kappa_{0}, \quad \mu=\mu_{0}=0 \, .
\end{equation}
With this assumption we find 
\begin{align*}
& U^{reg}\left(x\right)=\frac{-x\cos x+\sin x}{x^{2}}
\\
& R^{reg}\left(x\right)=\frac{4\sqrt{\pi/3}\sqrt{G\kappa_{0}}\rho_{0}\sin x}{x}
\\
& U^{f}\left(x\right)=-\frac{3x\sqrt{3\kappa_{0}}\Omega^{2}}{32G^{3/2}\pi^{3/2}\rho_{0}^{2}}
\\
& R^{f}\left(x\right)=-\frac{9\kappa_{0}\Omega^{2}}{8\pi G\rho_{0}},
\end{align*}
where $x=4\sqrt{ G\pi / (3\kappa_{0}) } \, r \rho_{0}  $.


\bsp
\label{lastpage}
\end{document}